# Cooperative Information Sharing to Improve Distributed Learning in Multi-Agent Systems

**Partha S. Dutta**                                          PSA01R@ECS.SOTON.AC.UK
**Nicholas R. Jennings**                                        NRJ@ECS.SOTON.AC.UK
**Luc Moreau**                                          L.MOREAU@ECS.SOTON.AC.UK
*School of Electronics and Computer Science,*
*University of Southampton,*
*Highfield, Southampton SO 17 1BJ, UK*

## Abstract

Effective coordination of agents' actions in partially-observable domains is a major challenge of multi-agent systems research. To address this, many researchers have developed techniques that allow the agents to make decisions based on *estimates* of the states and actions of other agents that are typically learnt using some form of machine learning algorithm. Nevertheless, many of these approaches fail to provide an actual means by which the necessary information is made available so that the estimates can be learnt. To this end, we argue that cooperative communication of state information between agents is one such mechanism. However, in a dynamically changing environment, the accuracy and timeliness of this communicated information determine the fidelity of the learned estimates and the usefulness of the actions taken based on these. Given this, we propose a novel information-sharing protocol, *post-task-completion sharing*, for the distribution of state information. We then show, through a formal analysis, the improvement in the quality of estimates produced using our strategy over the widely used protocol of sharing information between nearest neighbours. Moreover, communication heuristics designed around our information-sharing principle are subjected to empirical evaluation along with other benchmark strategies (including Littman's Q-routing and Stone's TPOT-RL) in a simulated call-routing application. These studies, conducted across a range of environmental settings, show that, compared to the different benchmarks used, our strategy generates an improvement of up to 60% in the call connection rate; of more than 1000% in the ability to connect long-distance calls; and incurs as low as 0.25 of the message overhead.

## 1. Introduction

A central challenge in multi-agent systems (MAS) research is to design mechanisms for co-ordinating agents that have partial, possibly mutually inconsistent, and inaccurate views of the system so that they can generate consistent solutions to complex, distributed problems. In such settings, the problem solving steps of one agent can influence those of another where they act on a common overall problem or use a set of sharable resources. Thus, to coordinate successfully, the agents need to *cooperate* by assisting each other to make better choices about the actions they take.[1] This cooperation is made more difficult because the individual agents usually have restricted capability in performing expensive computational activities (due to limited memory, CPU cycles, communication bandwidth, and/or communication latency) and because the target environment is usually characterised by continuous and unpredictable changes which, in turn, necessitate continuous adaptation of the problem solving process by the agents.

---

1. A different philosophy studied by MAS researchers is that of using *competitive* agents to find solutions to distributed problems (Takahashi & Tanaka, 2003; Walsh & Wellman, 2003). Here, however, we specifically focus on cooperative agents and thus we are firmly in the realms of cooperative distributed problem solving systems (Allsopp, Beautement, Bradshaw, Durfee, Kirton, Knoblock, Suri, Tate, & Thompson, 2002; Lesser & Erman, 1988; Mailler, Lesser, & Horling, 2003).

To overcome these problems and to coordinate effectively, the agents need a mechanism to act adaptively such that a consistent overall solution is generated. In this context, multi-agent coordination is typically based on techniques of modelling the states of other agents and enabling an agent to take actions based on these models (Dutta, Moreau, & Jennings, 2003; Dutta & Sen, 2003) (see section 2 for more details). However, given that the agents can only directly observe a limited subset of the system, they need to be provided with some information about the unobservable states to generate such models. To achieve this, we believe the agents should share some of their knowledge about their own locally observed states (at a suitable level of abstraction). This knowledge can then be used by the receiving agents to take more informed actions for better coordination. Note that in this approach, the agents cooperate by voluntarily distributing information in the system to facilitate the problem solving process. Nevertheless, practical resource bounds such as limited bandwidth and latency prohibit the use of exhaustive communication so that all agents could be made aware of the status of all other agents at all times. Thus the communication must be selective and aim to communicate the minimal amount of information that is necessary for effective coordination.

Given such constraints, it is not practical (nor possible in most cases) to generate models of all agents in the system that are accurate and up-to-date. Nevertheless, the agents can usually generate *estimates* of the un-observed states to take coordination decisions. Further, in dynamically changing systems, the agents should have a way of updating these estimates to *adapt* their problem solving decisions with the changing environment for generating quality solutions.

Now, in many applications, reinforcement learning (Sutton & Barto, 1998) (RL) has been successfully used (Ernst, Glavic, & Wehenkel, 2004; Mahadevan, Marchalleck, Das, & Gosavi, 1997; Tong, 2002) to generate such adaptive estimates in dynamic environments (see section 2 for a discussion of other alternatives). RL uses prior experience of performing tasks to develop a model of the environment. Specifically, a reinforcement learning agent receives a certain "reward" for taking an action in a given state, that acts as feedback to indicate the quality of performance against the context defined by the state-action pair. Using such rewards, RL is capable of incrementally generating robust estimates of the outcomes of different actions in different states. Such estimates provide the agent with a generic (independent of the problem) mathematical reasoning mechanism to take adaptive decisions in dynamic environments. In particular, the Q-learning (Watkins & Dayan, 1992) variant of RL is widely used because it allows estimates to be learnt without having prior knowledge about the system dynamics. Using Q-learning, with suitable training of the agent, where it is allowed to repeatedly take different actions in different states, the correct environment model can be learned from rewards. However, the assumptions underlying this result are that the agent is able to *observe all environmental states* and *receive all rewards accurately* for any action taken at any state. But, in practical MASs, this assumption is impossible to realise because an agent can only observe a part of the complete system. This implies that it can only perceive the result of its actions within its local environment and that it may not be able to observe immediately and correctly the actions (or, their effects) taken by other agents.

To overcome the above-mentioned limitations, we have developed an effective and efficient communication protocol that a set of cooperative agents can use to learn the estimates of unobserved states in a dynamic environment. In particular, we present a novel principle of *post-task-completion* (PTC) *information sharing*. In this, agents take actions for solving a given task using their *current* estimates of the system states and then distribute their local state information to one another only *after* the task is completed. No communication is assumed during the period of task processing. Then, information is shared between those agents who cooperate to complete the task. Upon receiving this information, the agents, subsequently, update their previous estimates of the states of the other agents.

This protocol is completely generic since it is not developed based on any domain or problem-specific assumptions (see analysis of section 5). However, specific instances of PTC can be implemented for a given problem domain. For example, in section 4, we describe such an instance of PTC implemented in a call routing problem where the agents attempt to estimate available bandwidth on nodes.[2] The fact that PTC is not based on any domain-specific assumptions implies that it can be used in cooperative multi-agent problems other than call routing and to verify this, we are currently studying its applicability in a distributed fault detection application with promising initial results (Dutta, Jennings, & Moreau, 2005).

The PTC protocol is distinct from the relatively standard approach of updating estimates using the information from only nearest neighbours (hereafter, referred to as NN) while processing a task which forms the basis of a family of routing protocols (Hedrick, 1988; Tanenbaum, 2003).[3] To emphasise this fact, we choose to compare the quality of estimates learned using the PTC principle against that of NN. Specifically, the NN protocol allows information to be shared between direct neighbours only, whereas our protocol allows information to be shared between a cooperative group of agents. Furthermore, our protocol ensures more timely communication which, therefore, leads to more up-to-date estimates than NN (section 5 establishes these formally).

There has also been other research (Shen, Lesser, & Carver, 2003; Xiang, 1996; Xuan, Lesser, & Zilberstein, 2001) that has studied how sharing information between agents can aid cooperative problem solving. Typically, these approaches treat communication as a distinct part of an agent's overall decision-making problem and show how its incorporation aids in solving the latter (section 2 has more details). But, there has been little in the way of a systematic study of developing a communication protocol that is both practically applicable (in terms of it being based on realistic assumptions) and effective (in terms of improving performance) in real-life MAS. Our work, on the other hand, investigates a specific communication protocol that has both the above desirable characteristics.

Against this background, in this paper, we evaluate PTC using two main approaches. First, by a mathematical analysis we demonstrate that the estimates generated by our approach are indeed more up-to-date than NN. Second, the effectiveness of our protocol is measured using a simulated distributed resource allocation problem.[4] In particular, we use a simulated wireless telephone network where the agents have to allocate bandwidth to connect calls. Communication heuristics, based upon the PTC principle, are devised to be used by the agents in this domain (see section 6 for a discussion). The performance of these are compared against two well-known algorithms used for network routing: Boyan and Littman's Q-routing (Boyan & Littman, 1993) and Stone and Veloso's Team Partitioned Opaque Transition Reinforcement Learning (TPOT-RL) (Stone & Veloso, 1999). The former is chosen because it is one of the most widely used benchmarks in learning-based network routing applications (Caro & Dorigo, 1998a; Peshkin & Savova, 2002; Stone, 2000). The latter has attracted attention more recently by being shown to be useful in a variety of domains (Stone, 2000). Hence, we believe these algorithms are reasonable benchmarks for empirical verification of PTC. Therefore, using both the formal analysis and empirical comparisons is sufficient to be able to firmly establish the merits of PTC.[5]

---

2. In this application, the "state" of an agent is the bandwidth availability of its node. More details follow in section 4.

3. In our example application domain, a node $i$'s "nearest" neighbours are those nodes that are within $i$'s transmission range, those with whom $i$ can directly communicate.

4. We choose resource allocation because it is a generic task domain widely used in practical MAS (Chaib-draa, 1995; Cockburn & Jennings, 1996; Jennings, Norman, & Faratin, 1998). Therefore, we believe it is a reasonable choice to test our information-sharing strategy. A preliminary empirical investigation of our information-sharing mechanism in this domain can be found in the work of Dutta, Dasmahapatra, Gunn, Jennings, and Moreau (2004).

Empirical studies have been conducted over a wide range of environmental settings by selecting different network topologies, network loads, and dynamically changing load patterns. These studies indicate very promising results for PTC. In particular, we observe substantial improvements in the rate of successfully connected calls (up to 60%) and the ability to route calls to distant destinations with high network load (more than 1000%); both achieved by incurring a much lower communication cost in terms of information message rate (as low as 0.25) by our PTC protocol compared to the benchmark strategies in the experimental settings used; all results being statistically significant at the 95% confidence level.

The following summarises our contributions towards advancing the state of the art:

- We argue for the use of information sharing based on realistic assumptions to improve cooperative distributed problem solving.

- We propose a communication protocol independent of problem-specific features — post-task-completion sharing — for generating good estimates that learning agents can use for better cooperation.

- We establish, using formal analysis, the advantage of the PTC protocol in generating more accurate estimates by ensuring a more timely distribution of information than the NN information-sharing protocol.

- We demonstrate the effectiveness of the PTC protocol using empirical studies in a representative multi-agent resource-allocation problem under a wide variety of environmental settings and against a range of other strategies.

The remainder of this paper is organised as follows. Section 2 discusses the general principles of multi-agent coordination and focuses on the specific approach of using machine learning techniques in this context. Also, the role of communication in these is analysed. In section 3, the characteristics of the cooperative multi-agent system on which we exemplify our research are outlined. It also contains a brief description of the network application that we simulate to empirically evaluate our communication principle. Section 4 presents a qualitative argument of the importance of using cooperative communication to improve learning. Also, how the PTC principle is implemented in the context of our example application is highlighted. Moreover, brief descriptions of the implementations of the benchmark algorithms (Q-routing and TPOT-RL) in this application are presented. Section 5 presents a formal analysis of the advantage of our strategy over the nearest-neighbour protocol in generating more accurate estimates. Subsequently, a detailed description of the simulation is provided in section 6. Section 7 describes the performance measures against which the various strategies are compared. It also analyses the results from our empirical studies. Finally, section 8 presents concluding remarks and identifies avenues of future work.

---

5. We also attempted to compare our algorithm with a global broadcast mechanism in which all agents issue a broadcast of their local state information whenever their states change. This mechanism was designed to verify whether system performance improves by transmitting all state-change information to all agents. Nevertheless, it is unsuitable to be used in a practical application due to its exorbitant message overhead. This was verified when we deployed it as a stand-alone application on a single machine with a dual 2.2 GHz AMD Opteron processor and 2GB memory in which it ran out of memory on the smallest topology and with the lightest load used in our experiments. However, a distributed implementation of the broadcast algorithm could be a matter of future study.

## 2. Related Work

In this section, we first review the major theoretical and empirical works on cooperative MAS that are developed around the theme of generating reliable estimates of unobserved states from limited interactions and adapting decisions in response to dynamic environments (section 2.1). Then, section 2.2 discusses cooperative MAS applications based on RL. As identified in section 1, RL makes adaptive decision-making possible without explicit domain knowledge or pre-defined rules of coordination. So it is used as the basic decision-making framework of our agents. This review specifically focuses on the use of communication in these applications and analyses the practical feasibility of the methods proposed. The shortcomings of these approaches are identified and the contributions of our research towards alleviating them is highlighted. Finally, section 2.3 discusses relevant literature in the area of network bandwidth estimation which is similar to our application domain and identifies how our learning-based approach differs from these.

### 2.1 Cooperative Multi-Agent Systems for Resource Allocation

In the following, we discuss the major research contributions in the area of cooperative MAS designed for resource allocation problems.

#### 2.1.1 Functionally Accurate Cooperation

The functionally accurate, cooperative (FA/C) approach advocates the exchange of partial, tentative solutions of local problems among agents to generate consistent partial solutions (a distributed speech recogniser application, based on this concept, is developed by Lesser and Erman 1988). In turn, this helps to generate predictive information about future partial solutions that furthers the build-up of a consistent global solution. However, the cost to obtain complete and up-to-date information to build a completely consistent solution can be prohibitively large because of communication delays. Hence, in such situations, it is more practical (cost effective) to achieve a global solution that may have a tolerable degree of inconsistency via the timely exchange of partial, tentative solutions. Thus FA/C is based on the use of communication to generate consistent estimates of the global problem solving scenario, as identified as a key requirement in section 1. Nevertheless, FA/C only discusses "what" is to be done, viz., agents should cooperate with partial solutions to reach an acceptable solution quality. It does not provide a recipe of "how" it could be achieved. Our work, on the other hand, advocates a specific communication strategy for the agents to improve the learning of state estimates which, in turn, would improve the cooperative problem solving. Moreover, we provide both a formal analysis and empirical results to justify the benefits of our strategy, something which the FA/C approach does not do.

### 2.1.2 ORGANISATIONAL STRUCTURING

Some researchers have incorporated organisational structures — patterns of information and control relationships that exist between the agents and the distribution of problem-solving abilities among them — into the agent models (Carley & Gasser, 1999). The idea is that such structures give an agent a broad, high-level knowledge about how the system solves problems, the roles that agents play, its own position in the network, and how they are connected. Imposing these structures, therefore, essentially, resolves the requirement (see section 1) of maintaining high-fidelity estimates of the portions of a system that the agents cannot directly monitor. This is analogous to the way human organisations are formed to solve complex tasks that are beyond the capability of "rationally bounded" (March & Simon, 1958) individuals. In our work, the agents do not follow pre-defined structures of roles and relationships. Instead, they learn, via cooperative communication, the estimates of the agents' states so that the dependencies can be inferred and the optimal actions can be taken. Thus, our cooperation model is applicable across domains without requiring explicit organisation structures to be specified.

### 2.1.3 SOPHISTICATED LOCAL CONTROL

The sophisticated local control methodology (for example, the *partial global planning* (PGP) approach Durfee & Lesser, 1991) advocates that the cooperating agents should reason about how to exchange information to resolve inconsistencies, whom to interact with to improve cooperation, what information exchange can achieve that objective, and the like.

Now, in PGP, the agents form contracts, plan their actions and interactions, negotiate over plans, use organisational information to guide their planning and problem-solving decisions, tolerate inconsistent views, and converge on acceptable network performances in dynamic environmental conditions. In this model, each agent maintains its own set of PGP's — a set of local plans that represent the agent's view of the global problem solving situation. They are updated by the exchange of local, partial plans among agents and reflect the most recent network scenario in terms of achieving the global solution. Hence, this methodology addresses the requirement (see section 1) of state estimation using cooperative communication.

Elaborating on the PGP approach, the *TAEMS* (Task Analysis, Environment Modelling, and Simulation) framework (Decker, 1995a, 1995b) was developed to model the impact that the characteristics of a task environment can have on coordination. Using *TAEMS*, coordination is achieved by three broad areas of agent behaviour: how and when to communicate and construct non-local views of the current problem solving situation; how and when to exchange the partial results of problem solving; how and when to make/break commitments made to other agents about what results will be available and when. The generalised partial global planning (GPGP) consists of a group of coordination mechanisms based on the above broad behavioural types (Decker & Lesser, 1995). Depending on the characteristics of the task environment, agents select the appropriate coordination mechanism. Unlike PGP, however, GPGP distinguishes local scheduling of an agent from its coordination activities: the coordination mechanisms provide an agent non-local views of the problem and the local scheduler creates plans (including both local actions and non-local effects via such actions) to improve global system-wide utility by using information from the coordination mechanisms. Thus, GPGP is developed around the principle of selecting actions based on estimates of the non-local states, as identified in section 1.

However, both PGP and GPGP employ, albeit flexibly, a set of predetermined coordination mechanisms. Such preplanned coordination can prove to be inadequate against all sorts of contingencies that can occur in domains where agents maintain incomplete, incorrect views of the world state (which change non-deterministically) and may even fail without prior indication. In contrast, our approach of learning to map the agents' views of the world states to the selection of actions which would guarantee the improvement of the global system performance, requires no such pre-specified coordination rules.

### 2.1.4 TEAMWORK BASED ON JOINT INTENTIONS

Probably the most comprehensive cooperative MAS framework existing in current literature is STEAM (Tambe, 1997) (see also Jennings, 1995; Rich & Sidner, 1997). It is developed around the principles of the joint intentions theory (Cohen & Levesque, 1991) and joint commitments (Jennings, 1993). To coordinate, the agents maintain a "joint persistent goal" (JPG) that the team is jointly committed to for doing some team activity, while mutually believing that they are doing it. Agents in STEAM arrive at a JPG by exchanging speech acts: "request", that they use to announce their individual partial commitments about attaining the global goal, and "confirm", which establishes that an agent has the same partial commitment to the one who made the "request". Further, STEAM borrows principles from the "shared plans" model (Grosz & Kraus, 1996) to ensure team coherence so that all team members follow a common solution path.

Although STEAM provides a principled framework for reasoning about coordination in teamwork, achieving a joint belief in large systems of widely distributed agents is, in most cases, likely to be a performance bottleneck rather than an advantage because of the excessive communication required to achieve it. This is especially true in environments where the agents have to re-plan the task execution process in response to environmental changes. Although STEAM treats this issue by using a replanning protocol, it requires the re-establishment of joint commitments among all the team members. A significant amount of communication overhead (message flow and delay) might be incurred before this is achieved which can degrade the quality of service significantly in some applications. Thus it is important to have a cooperation model that allows the agents to continue solving the overall task without requiring them to establish a system-wide commitment whenever replanning occurs. Our learning-based cooperation model has this advantage.

## 2.2 Learning-Based Cooperative Multi-Agent Systems

Now we review some key applications that use machine learning techniques for solving multi-agent cooperation problems.

### 2.2.1 Q-ROUTING IN DYNAMIC NETWORKS

RL is applied to the problem of cooperative distributed problem solving in a seminal piece of work by Boyan and Littman (Boyan & Littman, 1993) where they solve a network packet routing problem. In their paper, they model each communication node on an irregular $6 \times 6$ grid as a reinforcement learner who maintains estimates of the delays in routing packets to different destinations. To route a packet to a given destination, an agent requests each of its neighbouring agents for their delay estimates for that destination node. Upon receiving the delay estimates of its neighbours, the requesting agent chooses the neighbour with the minimum delay estimate to forward the packet. It then updates, using standard Q-learning, its prior delay estimate for that destination with the estimate that it received from this neighbour.

The authors of this paper demonstrate, using empirical studies, that their approach enables the agents to learn better policies (in terms of choosing a neighbouring agent for routing to a given destination) than a hand-coded shortest path algorithm. The differences are more pronounced when the network load increases indicating that the learning algorithm is able to adapt routing decisions (the paths along which packets are routed) under dynamic network conditions. In addition, the authors test their algorithm with changes in the network topology (by manually breaking the links between certain nodes) and in the pattern of call traffic (changing different regions in the network where calls can originate and terminate). They demonstrate that their Q-routing algorithm successfully adapts to these changes and performs better than the deterministic shortest path algorithm.

In their work, therefore, Boyan and Littman have used a simple communication protocol to allow the agents to cooperatively share their own knowledge about the packet routing delays. Nevertheless, the communication protocol they have used only allows an agent to inform its immediate neighbour about its own knowledge. This method would incur long latency for information to reach agents further away. As a result, considering states change continuously, the information can become outdated by the time an agent receives it. Such outdated information would then be of little use to generate reliable estimates of the non-local states. In this context, therefore, it is envisaged that by allowing state-change information to be shared between the group of cooperating agents only after task completion, the agents can maintain more accurate estimates of their non-local states. This, in turn, can improve the overall performance of the cooperative MAS as achieved by Boyan and Littman (1993). Moreover, in their work, the agents update their Q-estimates with the *estimates* received from direct neighbours. Note that in so doing, one learner depends on the estimates learned by another. Thus, this approach (essentially, a TD(0)-type learning Sutton, 1988) has the potential pitfall that "bad" estimates are propagated due to the poor learning of one agent. PTC, on the other hand, advocates transmitting the *actual states* of individual agents to the others in a group. Thus, PTC is not envisaged to have the shortcomings of the approach of Boyan and Littman (1993). To verify this, we choose the Q-routing algorithm as one of our benchmarks for empirically evaluating the performance advantage of PTC.

### 2.2.2 Team Partitioned Opaque Transition Reinforcement Learning

The **T**eam **P**artitioned, **O**paque **T**ransition **R**einforcement **L**earning algorithm (TPOT-RL) (Stone, 2000) has the objective to make the learning task easier in a MAS by reducing the state space dimensionality. It does this by mapping the state onto a limited number of *action-dependent* features. Analogous methods of state aggregation have been used in other reinforcement learning algorithms (e.g., McCallum, 1996; Singh, Jaakkola, & Jordan, 1995) to reduce the size of the learning task. However, TPOT-RL differs from these approaches because it emphasises deriving a set of small yet informative features for effective learning. More specifically, these features are used to represent the short term effect of actions that an agent may take. Thus, the agents learn the utility of selecting actions with respect to their own feature space. This is especially useful when the agents cannot observe or immediately influence the actions taken by other agents (such as in many practical multi-agent settings, in particular, network routing). That TPOT-RL is an effective algorithm is demonstrated by its successful application across multiple domains (Stone & Veloso, 1999).

The authors have evaluated the TPOT-RL algorithm in a simulated network routing environment. The action-dependent features in this case are the load levels of a node's adjacent links. The agents transmit their delay estimates along with a packet while routing the latter. Furthermore, these estimates, collected at the corresponding destination nodes, are distributed to the nodes who participated in the routing after fixed time intervals. Thus, TPOT-RL in fact uses communication to distribute information among the agents. Their empirical studies demonstrate that TPOT-RL outperforms (performance measure is average packet delivery time) the shortest path and Q-routing protocols when learning is done under switching traffic conditions — the algorithm is trained under conditions where the selection of packet sources and destinations are changed to form two different traffic patterns. Nevertheless, the following limitations are envisaged in this work. First, identifying action-dependent features from local observations only can lead to loss of information. This is because not all non-local state changes may be reflected in an agent's immediate state space but such information may be required by an agent to select actions. Thus, in such circumstances, explicit knowledge of the non-local states is necessary. Second, as a consequence of the above-mentioned problem, the fidelity of the derived estimates would deteriorate. This, in turn, would decrease the overall performance of the system. Third, in the work of Stone (2000), since the agents update their estimates based on others' estimates and not using the actual states, a similar shortcoming as identified in section 2.2.1 of learning bad estimates can occur. In contrast, our work attempts to alleviate these limitations by distributing the actual node states to keep the agents informed of the non-local states. Finally, in TPOT-RL, information is distributed at regular intervals (Stone, 2000), however, a formal way of specifying this interval is not prescribed. This is an arbitrary scheme which can result in large latencies in information reaching target nodes. Hence, estimates generated based on such information may not be up-to-date. PTC, on the other hand, distributes information immediately after task completion, thus, attempting to minimise the latency. Again, because of its claimed effectiveness and broad applicability, we choose TPOT-RL as the second benchmark for empirical evaluation against PTC.

### 2.2.3 POLICY GRADIENT SEARCH

Another approach of using RL for cooperative distributed problem solving is that of policy gradient search (Sutton, McAllester, Singh, & Mansour, 2000). A *policy*, in a RL context, is a mapping from a state on to an action. The policy is thus a function of a set of *parameters* which are variables defining the local state and, hence, influencing the action selection. A policy gradient search is a mechanism that tries to optimise the parameter values such that the average long-term reward of the learners is maximised. For example, in a network routing problem, these parameters can be the destination of the packet and the outgoing link a router (agent) selects for that destination (note these parameters are locally observable to an agent) while the reward is a measure of utility (or, sometimes the negative cost) that an action achieves given the parameter values. In the network routing domain, the reward can be the negative trip time for a packet to reach its destination node. In the policy gradient approach, it is assumed that each agent (the individual learners) receives the reward of all actions taken by all agents at every time step.[6] It is only this reward information that is globally known by the agents. Thus the policy gradient algorithm is model free — independent of domain models and knowledge about others' states and actions. Individual agents adjust their policy parameters in the direction of the gradient of the average reward that they compute using the global reward information — hence the term policy *gradient*. Therefore, communication of reward values is key to allow the learners to optimise the parameter values. However, the dependence on the global reward information to update the policy parameters can be a bottleneck in systems where the communication bandwidth is limited and there is a finite latency in messages to propagate (as in most practical systems). These constraints can lead to very slow responsiveness to environment changes in agents using the policy-gradient approach. Moreover, broadcasting rewards by all agents in highly dynamic environments (such as, networks experiencing heavy loads) can cause the network to completely saturate by the reward messages (as observed in our implementation of the global broadcast strategy, stated in section 1) resulting in a very inefficient system.

This method is used to build cooperative MAS by Williams (1992), Baxter and Bartlett (1999), and Peshkin and Savova (2002), among others. All of these works demonstrate that the policy-gradient search achieves reasonable performance (in terms of average routing delay) compared to other benchmark algorithms such as the shortest path algorithm. However, in the works of both Williams (1992) and Baxter and Bartlett (1999), an exceedingly large amount of time is required for the learners to converge. This is because the learners need the global reward information to update their policy parameters which results in a slow optimisation of the parameters. This puts a restriction on the applicability of this approach to build practical systems. A similar limitation is likely in the work of Peshkin and Savova (2002) although the authors do not provide these results. In addition, with the high communication overhead incurred, this approach is likely to be unsuitable for implementing practical MASs.

In contrast to the uninhibited communication required in the policy gradient search approach, in our work, communication is used as a controlled strategy to inform the agents about the portions of the global state that are relevant to their action selection. Thus, we believe our work offers a superior practical solution to the policy gradient approach.

### 2.2.4 Communication Decisions in Multi-Agent Coordination

Xuan, Lesser, and Zilberstein (2001) advocate that communication decisions are integral to an agent's decision to coordinate in a cooperative, distributed MAS. The authors consider that each agent solves a local Markov Decision Process (MDP) (Feinberg & Schwartz, 2001) that generates both a communication action and a state-changing action at every decision sub-stage. The agents are only given local observability (i.e., they cannot observe the states of other agents). However, they can observe the communication actions of other agents. The important reason for introducing a communication decision in an agent's local MDP, as argued in this paper, is that communication incurs cost. Hence, an agent should employ reasoning to decide when communication is required such that the overall utility earned from the agent's decisions is maximised. In this context, this work extends the theoretical analysis of multi-agent MDP of Boutilier (1999) where the agents are assumed to have global state information. Specifically, the authors propose two simple heuristics to generating communication decisions that aim to reduce the computational complexity of solving the full MDP to generate the optimal global policy.

---

6. More realistically, each agent may broadcast the reward it receives to all other agents. Hence, the agents may receive the reward signals from the entire system with some delay.

As we are interested in studying the impact of communication on the performance of a cooperative MAS, the work of Xuan et al. (2001) is related to our research. However, while they analyse *whether* communication is necessary at any stage of an agent's decision, we consider communication to be inevitable. We analyse, both quantitatively and empirically, the impact of a specific information-sharing protocol, PTC, on the performance of a MAS. Our work differs from that of Xuan et al. (2001) in the following additional ways. Firstly, in their work, the agents are assumed to iterate through a sequence of communicate and act stages *synchronously*. We adopt a more generic approach of completely asynchronous behaviour. Secondly, they assume to be instantaneous communication; thus, information sent by an agent is received by another immediately. On the contrary, we consider a more realistic scenario where there is a finite delay associated with communication. Finally, the communication heuristics that they propose are based on each agent individually monitoring their own progress towards achieving a commonly agreed upon goal. In our work, we consider fundamentally distributed task processing where an agent can take a local action and the actions of multiple agents together complete a task (more on this in section 3). Hence, in our work, individual agents cannot monitor the progress of a task execution process towards completion; they are only capable of taking their local actions using estimates of the unobserved states.

### 2.2.5 OTHER MACHINE LEARNING ALGORITHMS

The motivation for PTC initially may appear to be similar to conventional supervised learning (SL) (Widrow & Hoff, 1960), where the actual outcome of a multi-stage prediction problem is fed back to the individual learners (predictors). However, they differ on the following critical issues:

- In SL, only the final outcome (such as whether a prediction was "correct") acts as the source of information for the learners to update their prediction algorithm. In contrast, PTC allows a learner to gain knowledge about how the states of other agents in the cooperative group change along with the outcome.

- In SL, an agent typically learns a mapping from its own actions onto the outcomes of a multi-stage prediction problem. PTC, however, allows an agent to consider the impact of the states of other agents on the final outcome of the task.

- Typically, SL is used for prediction in stationary environments. On the other hand, we evaluate the competence of PTC in maintaining high-fidelity estimates in fundamentally dynamic and uncertain environments.

Also, PTC is distinct from the approach of using eligibility traces in which a learner is provided with the knowledge of the entire sequence of state transitions after a complete training episode (i.e., after starting from the start state and reaching the goal state) (Mitchell, 1997). In the latter, an agent, upon reaching the goal state after executing a series of actions, updates in reverse order (i.e., starting from the goal state and moving to the starting state) its Q-estimates for each state transition. In a practical MAS, however, it is not possible for a single agent to observe all transitions occurring during a task processing episode as assumed in the approach using eligibility traces. Further, in a large and complex MAS, the computational load incurred by a single agent attempting to take decisions based on its knowledge of all state-transitions sequences of every task would be too prohibitive to be realised in practice. In this situation, therefore, our research contributes towards developing a practical and effective means of distributing state information to improve learning. In so doing, it removes the requirement of an agent having to maintain the entire chain of state transitions in its memory. Rather, it allows the agents to acquire an overall picture of the state-changes in the cooperative group that perform a task, which, in turn, allows them to take decisions for effective coordination.

## 2.3 Network Bandwidth Estimation

Network bandwidth availability directly impacts the performance of networked applications such as web services, peer-to-peer systems, and mobile networks. Therefore, effective estimation and prediction of bandwidth availability have attracted considerable attention in the networks community. Our application of PTC to do bandwidth estimation and routing, therefore, bears close resemblance to this line of research. In this paper, we use Q-learning (Watkins & Dayan, 1992) for bandwidth estimation because it generates robust and flexible estimates from observations. Therefore, to generate the estimate, it needs to observe the bandwidth availability pattern. In network bandwidth estimation, such knowledge is harnessed by either *active* or *passive* measurements. The former is done by injecting perturbation traffic into the network and then assessing the states based on the statistical characteristics of this traffic (Jain & Dovrolis, 2002). However, in applications such as limited-bandwidth ad-hoc networks, such perturbation traffic wastefully consumes valuable bandwidth. A related approach adopted in the agent's community is that of using mobile agents or "ants" to harness traffic conditions across the networks and update node routing tables (Caro & Dorigo, 1998a, 1998b). In low-bandwidth networks, the introduction of such extraneous agents can still impact bandwidth usage. Besides, the security and privacy problems associated with the use of mobile agents have made their applicability in real-life systems debatable. Alternatively, passive measurements are done by offline analysis of actual traffic traces (Lai & Baker, 1999; Ribeiro, Coates, Riedi, Sarvotham, Hendricks, & Baraniuk, 2000). However, in our case, we require estimation to be online so that we can cope with the dynamic conditions during the operational period of the system. Therefore, PTC generates bandwidth estimates by using information disseminated by the cooperative agents *while* they route calls. Further, the passive measurement approaches typically estimate bandwidth by assuming a network model that can affect the estimation fidelity. For instance, in the work of Ribeiro et al. (2000), a network path is modelled as a single queue which disregards the effects of variable queueing delays along that path on the estimate. By choosing Q-learning, we aim to develop a statistical model of bandwidth availability by continuous monitoring without getting constrained by any predefined models.

## 3. Task Domain Characteristics

In most MAS, complex tasks are performed by groups of agents. Generally speaking, the problem solving activities of such agents may be executed in parallel to generate the final solution. But, in many cases, constraints over processing different, but related, parts of a task by different agents may require the task execution process to be partially sequentialised with appropriate scheduling between parallel executions (Maley, 1988; Parunak, Baker, & Clark, 2001). As a simple example, we can consider a car factory where the units manufacturing the different parts of a car can run in parallel (satisfying mutual compatibility) while the final assembly comes into play only after all the parts are correctly manufactured (Jennings & Bussmann, 2003). Alternatively, the nature of a task can enforce a strictly sequential processing. For example, in a distributed transportation system, delivery of goods between two points requires a set of cargo movements in sequence. In this paper, we exemplify the application of our PTC principle in a sequential domain. However, the choice of a sequential domain does not indicate a limitation of PTC; PTC is not based on any assumption of sequential task processing (section 4 has more details). Nevertheless, this is a reasonable choice since several key MAS applications such as sensor networks (Viswanathan & Varshney, 1997), supply chains (Denkena, Zwich, & Woelk, 2004), telecommunication bandwidth allocation (Minar, Kramer, & Maes, 1999), among others, feature, to different extents and at different levels of granularity, sequential task processing. Here, it should be clarified that although individual tasks are considered sequential, the entire multi-agent system would typically be performing multiple such sequential tasks simultaneously and asynchronously. Further, the individual agent activities in our example application are considered equivalent to allocating resources. This is because in many practical MAS applications, such as the ones mentioned above, the agents essentially allocate resources such as network bandwidth, processor cycles, etc. to complete tasks.

Specifically, in this research, we consider a wireless telephone network (TN) as a representative application in which to implement communication heuristics based on PTC and evaluate these empirically. Note, however, that this choice does not limit the applicability of our results. This is because this application has characteristics that are common to many real-world large-scale distributed systems and hence our solutions are also more generally applicable. These include: multiple agents situated in different portions of the system (a network node is modelled as an agent), agents having incomplete views of the activities and states of others, and agents having to coordinate their local (routing) decisions in order to successfully achieve a global task (routing a call from source to destination). Moreover, there is a clear measure of success which is proportional to the bandwidth usage efficiency, or, equivalently, to the total number of successfully routed calls.

The TN uses circuit-switched communication where node bandwidth has to be allocated end to end (from a call source to destination) to establish calls. Each routing node is treated as an agent. A node has a certain amount of bandwidth that can be allocated to a certain maximum number of calls simultaneously. Each agent can monitor the load (the amount of bandwidth allocated to calls) on its node and can communicate only with the nodes within its transmission range — the set of "neighbour" nodes. Calls of finite duration can originate from and be destined to any node. Calls originate (so, terminate) continuously. Therefore, the load on the nodes continue to vary with time. The objective of any such agent is to allocate bandwidth and forward a call to one of its neighbours such that the probability of the call getting routed via the least congested (with the maximum available bandwidth) path is maximised. Routing a call at a given time along the least congested path at that time ensures an efficient use of bandwidth, hence, increases the number of successfully routed calls in the system. The forwarding is based on the agent's estimate of the congestion levels across the network, hence, its estimate of the unobserved states. Hence, the task is completed by a sequence of such allocations and forwarding by multiple distributed agents. In this scenario, the agents continually process tasks (i.e., route calls from source to destination as new calls originate). We refer to the process of routing a call as a task processing *episode*. Section 5 uses this notion of episodic tasks to explain the difference between the estimates generated by PTC and the NN protocols. A detailed description of the simulation of various agent activities during such episodes is provided in section 6.

The next section expands on the PTC principle and outlines how information sharing based on this principle is designed in the TN domain. Furthermore, the implementations of the benchmark strategies, Q-routing and TPOT-RL are also explained.

## 4. Sharing Information to Improve Learning

To implement effective learning in dynamic environments by sharing local knowledge between agents, the communication strategy should satisfy the following criteria:

- **Time efficient distribution:** There is a latency associated with communication. Hence, the more timely the information that is communicated to an agent, the more likely the information is up-to-date.

- **Accuracy of information:** In continuously changing environments, it is impossible for all agents to remain synchronised with state changes at all times. Nevertheless, the more accurate the information received, the better.

Given these desiderata, here PTC is proposed as an effective strategy for distributing the local state information of agents.

**Definition 1** *Post-task-completion information sharing refers to the distribution of local state information between a group of cooperative agents, by way of a mechanism that depends on the allowed agent interactions, only after the completion of tasks undertaken by the agents.*

The motivation for using this scheme is to let an agent that participated in completing a collaborative task have an indication of the state changes of the other agents in the group that resulted from processing that task. Such information is then useful for making more informed decisions while processing any subsequent task. In a dynamic system, the world states change while the agents process a given task. Therefore, by delaying the transmission of information until the task is completed, this protocol ensures that all those agents who participated in the task completion process are informed about these state changes and how that affects the outcome of the task. In so doing, we hypothesise that by using this mechanism the agents would be able to distribute information in a time-efficient manner and learn reasonably accurate estimates, thereby satisfying the requirements identified above. The analysis presented in section 5 establishes our hypothesis by comparing the timeliness and estimate qualities of PTC against those of the NN protocol.

It is important to note that PTC is a general principle and specific instances of PTC can be implemented in a given problem. Our implementation of PTC in the TN domain, discussed in section 4.2, is just such an instance. Further, although we exemplify the applicability of PTC in a sequential domain, the protocol is not designed around any such assumption. Note that in applications where task processing between agents occurs independent of one another and in a purely parallel fashion, the objective of maintaining estimates of other agents' states becomes redundant. Rather, in such systems, a central task allocator that allocates sub-tasks to the parallely executing agents would be a more suitable approach. However, if the processing of a sub-task by an agent requires estimates of the others (due to some dependencies between them), PTC can be used to distribute information between these processors. Moreover, if the sub-tasks in such systems, in turn, require sequential processing, our information-sharing protocol can be used in that context. Therefore, we can argue that communication strategies based on the PTC principle can be suitably designed to be applicable at various levels of granularity in domains other than the one chosen in this paper.

In the remainder of this section, we first outline the basics of Q-learning. Subsequently, we describe how PTC is implemented in a TN to improve the distributed learning of bandwidth availability. The Q-routing and TPOT-RL implementations are also described in the same light.

## 4.1 Q-learning Basics

Q-learning is an algorithm to learn the optimal action an agent can take such that the discounted cumulative utility of a sequential decision problem is maximised. In general, RL achieves this task by assuming full knowledge of the reward generated for any action in any state and the resultant state after an action is taken. A formal description of RL requires that the environment be represented by a *4*-tuple: $\langle \mathcal{S}, \mathcal{A}, \mathcal{T}, \nabla \rangle$, where $\mathcal{S}$ is a finite set of states, $\mathcal{A}$ a finite set of agent actions, $\mathcal{T} : \mathcal{S} \times \mathcal{A} \to \mathcal{P}(\mathcal{S})$ a probability distribution $P$ over states generated by a given action taken in a given state, and $r : \mathcal{S} \times \mathcal{A} \to \mathcal{R}$ returns a scalar value (the reward) as a result of an action taken in a state. Now, the policy of a RL agent $\pi : \mathcal{S} \to \mathcal{A}$ is a mapping of state on to actions. Given a policy $\pi$, the *value function* $V^\pi(s_t)$ refers to the cumulative reward that the agent receives by following $\pi$ starting at state $s_t$ and all subsequent states $s_{t+i}, i \geq 1$. Thus,

$$V^\pi(s_t) = E \left[ \sum_{i=0}^{\infty} \gamma^i r_{t+i}^\pi \right]. \tag{1}$$

where $\gamma$ ($0 \leq \gamma < 1$) is a discount factor that determines the relative weights of immediate and delayed rewards, and $r^\pi$ refers to the reward received at each state transition following policy $\pi$. Note the expected value is used because $r$ typically defines a probability distribution over the outcomes generated by taking an action in a given state. In the above expression, the agent's "life" is assumed to be infinite, hence the sum is over an infinite sequence. The *optimal* policy $\pi^*$ is:

$$\pi^* = \underset{\pi}{\operatorname{argmax}} \, V^\pi(s), \forall s. \tag{2}$$

The action generated at any state $s$ by the optimal policy $\pi^*$ is:

$$\pi^*(s) = \underset{a}{\operatorname{argmax}} \, E \left[ r^\pi(s, a) + \gamma \sum_{s'} T(s, a, s') V^{\pi^*}(s') \right], \tag{3}$$

where $V^{\pi^*}$ represents the value function corresponding to the optimal policy $\pi^*$. Expression (3) indicates that the optimal policy can be acquired by learning the optimal value function for all states. Nevertheless, in so doing, complete knowledge about the $r$ and $T$ functions is necessary. However, in most practical domains, an accurate knowledge of these two functions is not possible.

Against this background, the usefulness of Q-learning is to learn the optimal policy *without* having to learn the optimal value function. The key to achieving this is the following

substitution:

$$Q^\pi(s,a) \equiv r^\pi(s,a) + \gamma \sum_{s'} T(s,a,s') V^{\pi^*}(s'). \tag{4}$$

The optimal policy can be calculated if an agent learns the $Q(s,a)$ values as,

$$\pi^*(s) = \operatorname*{argmax}_a Q(s,a), \tag{5}$$

and the optimal value function is computed as,

$$V^{\pi^*}(s) = \max_a Q(s,a). \tag{6}$$

A recursive definition of the $Q$ values is thus,

$$Q^\pi(s,a) \equiv r^\pi(s,a) + \gamma \sum_{s'} T(s,a,s') \max_{a'} Q(s',a'). \tag{7}$$

Without any knowledge of the reward and state transition functions, the true $Q$-values (shown in (7)) can be estimated through repeated training. In particular, the following training rule, due to Watkins and Dyan, allows the estimated values, $\hat{Q}$, to converge to their true values (Watkins & Dayan, 1992):

$$\hat{Q}_{n+1}(s,a) \leftarrow (1-\alpha)\hat{Q}_n(s,a) + \alpha[r(s,a) + \gamma \max_{a'} \hat{Q}_n(s',a')]. \tag{8}$$

Here, $s$ and $a$ are the state and action, respectively, updated for the $n+1^{th}$ training iteration, $\alpha$ ($0 \leq \alpha \leq 1$) generates a decaying weighted average of the current $\hat{Q}$-estimate and the revised value, and $s'$ is the resultant state after taking action $a$ in state $s$ according to the current policy. By adjusting (gradually reducing) the value of $\alpha$, the $\hat{Q}$-estimates can be demonstrated to converge to the actual $Q$-values (Watkins & Dayan, 1992). In this research, we let our agents use the training rule (8) to learn the $Q$-values.

## 4.2 Strategies for Sharing Information in the Cooperative Bandwidth Allocation Problem

Assuming a TN has a set of agents $\mathcal{A}$, we consider an arbitrary subset of $\mathcal{A}$, $\mathcal{N} = \{a_i \mid i = 1, ..., n\}$ (where $n$ is the cardinality of $\mathcal{N}$), that processes a given task at a given time. In so doing, an agent in $\mathcal{N}$ considers one of its neighbours to forward a call (refer to the discussion in section 3). This decision is based on the agent's Q-estimates. Specifically, we use a Q-table for each agent $a_i$ where an entry $Q_i(n,k)$ represents the expected utility of choosing neighbour $a_k$ when the call destination is $a_n$ (note the size of the Q-table for each agent is $\mid \mathcal{A} \mid \times \mid \mathcal{K} \mid$, where $\mathcal{K}$ is the set of neighbour agents). In particular, for a TN, we chose the Q-values to represent the *expected availability of free bandwidth (BW) channels* on nodes along the various paths from $a_k$ to $a_n$. Note, in this representation of the Q-function, $n$ is the *goal state* (the fact that the call has to be routed to $a_n$) of the agent and the *current state* is $i$ (the fact that the call is currently with $a_i$). So, effectively, the agent learns to forward a call along the path with the maximum available BW to reach the goal state from the current state. This representation has the advantage that all intermediate state transitions (the sequence of nodes that the call has to be routed through) are collapsed into one effective transition from the current state to the goal state. The Q-value, therefore, signifies the "effective utility" (in terms of the available bandwidth; the higher the value of which the better is the utility in terms of successfully routing a call) of selecting a given neighbour to reach the goal state. This Q-value is learned from the information distributed by the information-sharing strategies, described shortly. Note, a similar Q-function has been used previously by researchers studying adaptive routing using Q-learning (Boyan & Littman, 1993).

In the above context, an agent ($a_i$) who wants to route a call to a destination node ($a_n$) using a set of nodes that have the highest expected BW availability, chooses the neighbour $a'_k$ such that $a'_k = \arg\max_{a_k} Q_i(n, k)$, where the maximisation is over all neighbouring agents of $a_i$. In our empirical study, we use Boltzmann's exploration (Watkins, 1989), a standard scheme for probabilistically choosing a neighbour as opposed to this deterministic strategy. An instance of agent $a_1$'s request to its neighbour $a_2$ to forward a call towards the destination node $a_n$ is shown in Figure 1(a). Being cooperative, $a_2$ will accept the request if it has the available capacity. After forwarding the request, $a_1$ pre-allocates one unit of call channel bandwidth (Figure 1(b)) for the *partially connected* call until it is either successfully connected or dropped (as described shortly). The forwarding continues (Figure 1,(b)) until the destination node ($a_n$) is reached. At this point, a message is transmitted back along the route through which the call was routed to inform each agent that the call has actually *connected*. Each agent then allocates one unit of call channel bandwidth to complete a circuit from the source to destination (before this, the nodes had only pre-allocated bandwidth) (Figure 1(c)). Also, using this message, agents transmit their local state values to other agents on the route. Hence, those agents that cooperated on a task (routing the call) share among themselves their local state information after the task is completed (after the sequence of requests reach the destination node). In this way, therefore, the PTC principle (see definition 1) has been instantiated specifically in the TN domain. More details on this follows shortly.

However, the forwarding process stops if an agent is contacted that has no unallocated call channel bandwidth. Then, the agent transmits a message to inform those on the route to drop the partially connected call and deallocate the pre-allocated bandwidth (Figure 1(d)). In addition, the worst case setup time of a call is bounded by an upper limit for the time that the agents can continue with the forwarding process. After this time, the call is dropped if it has not connected. This time is equivalent to the maximum delay a caller would experience between dialling a number and hearing the ring tone. [7] Finally, if an agent, while forwarding the call, detects a cycle in the route taken by the sequence of requests it generates a message with a penalty and transmits it to the agents on the cycle. This penalty, an exponentially decreasing function with the distance of a node from the loop end, is subsequently used by those agents to update their estimates such that the occurrence of further cycles is reduced. Moreover, while penalising, the agents on the loop de-allocate the previously pre-allocated bandwidth since loops are redundant portions of a call path (for more details, see section 6.2).

In the preceding, we presented a broad description of the sequential bandwidth allocation task performed by the routing agents. In the following, we first explain how the PTC principle is implemented in this setting. Then, the implementations of the benchmarks in this domain are presented. A more detailed discussion of our system implementation is provided in section 6.

---

7. In case of the above two conditions, we do not use any backtracking to search for alternative paths. This keeps the routing protocol simple and makes the analysis of the system behaviour easy. Moreover, this simplification should not impact the overall conclusions of this research since inclusion of backtracking would impact all strategies equally.

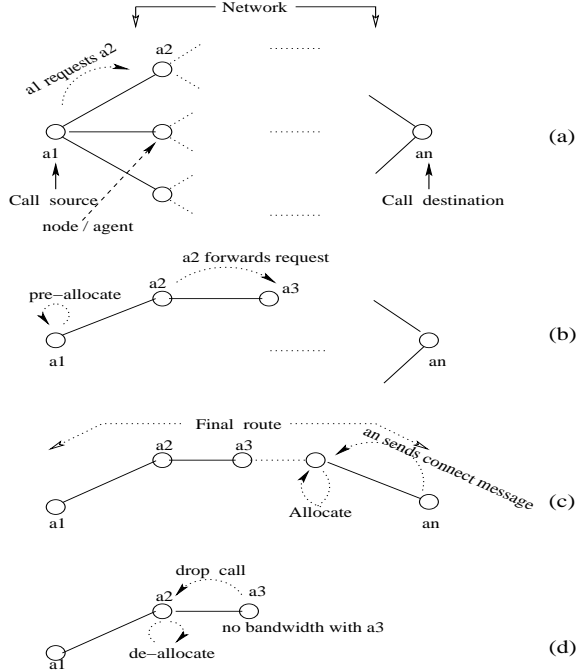

Figure 1: The call forwarding process

### 4.2.1 Post-Task-Completion Information Sharing

Consider (as described above) a call is routed from $a_1$ (source) to $a_n$ (destination) along the path $a_1...a_n$. In the instantiation of the PTC principle, agent $a_n$ starts communicating its own local state (in a TN a node's state is represented by the available BW units on that node) when the BW allocation process completes at time $t$ at $a_n$, thereby, establishing a complete circuit from source to destination. Agent $a_n$ communicates this information to $a_{n-1}$. Thus, $a_{n-1}$ updates its prior estimate of $a_n$'s available BW units $Q_{n-1}(n, n)$ with the new state information using the standard Q-update rule (equation 8): $Q_{n-1}(n, n) \leftarrow (1-\alpha)Q_{n-1}(n, n) + \alpha\, s(n, t)$, where $s(n, t)$ represents the local state of $a_n$ at time $t$ and is the "reward" for the Q-learner to update its prior Q-estimate. Subsequently, $a_{n-1}$ communicates to $a_{n-2}$ its own local state at time $t'$ $s(n-1, t')$ ($t' \neq t$, because of the latency in communication between neighbour agents) and the information it had received from $a_n$. Alternatively, it can use its own state information and that received from $a_n$ to communicate a summary information that captures the overall state of the path being used to route the call. Section 6.3 outlines two heuristics of doing this.[8] Agent $a_{n-2}$ similarly updates its prior estimate of the bandwidth availability $Q_{n-2}(n, n-1)$ using the information received from $a_{n-1}$. This procedure of distributing their own and the previously received state information continues until the source agent (here, $a_1$) is reached. Typically, the information about the states of multiple nodes is used to generate a summary estimate of the state of the downstream route as described in section 6.3. The distinction between local state (an agent's own state) and non-local state (another agent's state) is lost in aggregating information of multiple nodes to maintain a summary of call routes. However, this is not a problem in a TN since the knowledge of path bandwidth availability is sufficient for an agent to take effective routing decisions (selecting a subsequent node to forward a call request). On the other hand, maintaining the information of individual nodes would necessitate each agent solving a computationally expensive least-cost-path problem before every routing decision. However, in a different domain, it is entirely possible for an agent to maintain separate state estimates of other agents in the cooperative group while using the same PTC principle.

---

8. Such summarisation of state information is separate from the basic PTC principle which simply states that information is shared between a cooperating group after task completion.

The realisation of PTC in the TN domain allows one agent at a time to communicate information to its immediate neighbour. This is because: (i) one agent (in a TN, the destination node) detects the completion of a task, which, subsequently, starts transmitting state information; (ii) an agent can communicate to its immediate neighbour only; and (iii) state information transmission follows a sequence (from the destination node towards the source) due to the previous two reasons. However, in a different multi-agent domain, a realisation of PTC may involve multiple agents sharing information at the same time depending on the interactions possible between agents.

In this particular instantiation of PTC, only those nodes who participated in routing a call share information. So, the state information of the other nodes in the system is not distributed. But, the decision of which agents should be informed about which system states is a separate problem. In this paper we advocate PTC as a specification for distributing information among those agents who cooperate on a task after task completion. Therefore, distributing information among the routing nodes follows this specification. Nevertheless, in our ongoing research, we are investigating the problem of how we can determine which agents within a cooperative group should be notified of a certain piece of state information after task completion and the consequence of such selective distribution.

The communicated state information acts as the "reward" for Q-learning. Therefore, the accuracy and the timeliness of this information is critical in determining the quality of the Q-estimates. This, in turn, directly impacts the effectiveness of an agent's decision to route calls.

### 4.2.2 Q-ROUTING

The Q-routing algorithm is based on the NN protocol (as discussed in section 2.2.1). In Q-routing (hereafter, referred to as QR), an agent $a_i$, after forwarding a call to neighbour $a_{i+1}$, receives the latter's current best estimate of the BW availability to reach destination $a_n$. Thus, neighbour $a_{i+1}$ informs $a_i$ with $\hat{Q}_{i+1} = \min(s_{i+1}, \max_{a_{i+2}} Q_{i+1}(n, i+2))$, where the maximisation is done over all neighbours $a_{i+2}$ of $a_{i+1}$. Since, on a given path in a TN, the node with the minimum bandwidth availability determines the maximum number of calls that can be placed via that path, $a_{i+1}$ determines the minimum of its own bandwidth availability (its "state", denoted by $s_{i+1}$) and its estimate of the subsequent path. Agent $a_i$, upon receiving this estimate, updates its prior estimate $Q_i(n, i+1)$ as: $Q_i(n, i+1) \leftarrow (1-\alpha)Q_i(n, i+1) + \alpha \hat{Q}_{i+1}$. This process of asking neighbours and receiving the latter's estimates continues until the destination is reached (see Figure 1(c)) or the forwarding process is terminated (see Figure 1(d)). The way the Q-estimates are updated in QR is similar to the update rule used in PTC as shown in section 4.2.1. The difference, however, is in the reward: whereas in QR, the reward is the estimate of the immediate neighbour, in PTC, it is a summary of the actual state information of all subsequent path agents.

### 4.2.3 TPOT-RL

TPOT-RL is implemented in our TN domain following the description given by Stone (2000). The characteristic features of TPOT-RL (see Stone, 2000 for more details) are implemented as follows: (i) a partitioning function and (ii) an action-dependent feature function (where the `activity-window` parameter is chosen as 100, and the usage-threshold for a link connecting to a neighbour is set to 5.0 — half the maximum bandwidth capacity of a node, defined in section 7.2), both identical to those of Stone (2000); and (iii) the reward `update-interval` is set to 100. Section 6.2 explains in more detail how these parameters are used by the agents to learn estimates and route calls. While forwarding a call, a node using TPOT-RL transmits to the subsequent node its current estimate about the bandwidth availability to reach the given call destination. Each agent also records the amount of bandwidth usage on each of the links connecting to its neighbours. Its Q-values estimate the bandwidth availability to reach a given destination via a given neighbour for the given link usage level, monitored over the past `activity-window` time steps. A call-forwarding decision is taken by using a Boltzmann exploration over the Q-estimates. Reward distribution occurs in TPOT-RL every `update-interval` time steps. More specifically, for all calls that are successfully connected, the corresponding destination nodes accumulate the information that was transmitted by the forwarding nodes along the call paths. Then, after every `update-interval` time steps, these destination nodes start sending the accumulated information back along the corresponding call paths. An agent located along such a call route, updates its Q-values after receiving this information. So, a node $a_k$ along the path $a_1, \cdots, a_k, \cdots, a_n$, gets the *estimates* (as opposed to the actual node states) of its subsequent nodes $a_{k+1}, \cdots, a_n$, using which it updates its Q-value. Note that we have used aggregation of the information received from subsequent agents similar to that in PTC (section 4.2.1).

## 5. The Advantage of the PTC Sharing Principle

In this section, we present a formal analysis to explain the advantage of our information sharing model in generating better learning than the NN protocol. While this establishes the benefits of PTC on a theoretical ground, it also provides an explanation for the performance improvements observed in our empirical studies. In the following, we first state our assumptions and notations that will be used in later discussions. The rest of this section develops a formal representation of the timeliness of distributing information by both strategies. This representation is then used to compare the accuracies of the non-local state information in each case.

### 5.1 Basic Assumptions and Notations

In section 3, it was stated that we consider task episodes that require the sequential participation of the appropriate agents for successful completion. In this context, our analysis focuses on a particular set of agents $\mathcal{N} = \{a_1, ..., a_n\}$ where agent $a_1$ initiates the task execution process by receiving a new task. Also, we assume, without loss of generality, that the order in which the agents process tasks is: $a_1 \rightarrow \cdots \rightarrow a_n$. This assumption implies that, in this task processing instance, $a_1$ uses its knowledge of the states of other agents and selects $a_2$ to forward the task, $a_2$ similarly selects $a_3$, and so on until $a_{n-1}$ selects $a_n$ which is the agent at which the task processing is completed (this was described in section 4.2 in the context of a TN). Note that $\mathcal{N}$ represents one possible set of agents that can complete the task (equivalently, in a network, there can be multiple routes through which a call can be routed to the destination). The state estimates that the agents use to select the subsequent agent are generated using communicated information of the unobserved states via the particular communication strategy used by the agents (PTC or NN). Thus, focusing on one particular set (in this case, $\mathcal{N}$) simplifies the analysis of how a communication strategy affects the accuracy of a given agent's knowledge of the agents in that set. Since $\mathcal{N}$ is arbitrarily selected, it is equivalent to selecting any other set of agents. Therefore, the results of our analysis do not depend on which $\mathcal{N}$ is chosen.

Further, we consider that the agents process tasks that are generated continuously (as per section 3). To this end, let the symbol $t_c$ denote the fixed time after which successive tasks are generated. Such an assumption of periodicity in the task environment is made to simplify our analysis. Subsequently, in section 5.6, we show that even under more general, non-periodic environments, the same general conclusions hold.

In the sequence of agents that jointly participate in executing the task, there is the notion of a "subsequent" agent (and, for that matter, a "preceding" agent) for any agent except the last (the first). We represent the agent subsequent to an agent $a_i$ by the identifier $a_{i+1}$ and the one preceding $a_i$ by $a_{i-1}$.

The agent state, as per section 4.2, is represented by a *real-valued* function $s$. For example, in a network, this can represent the load level, or, equivalently, the fraction of the total bandwidth used on a node. Also as discussed in section 4.2, an agent learns these agent states, using the communicated information from other agents, to decide which subsequent agent to choose. The *actual* state of agent $a_i$ (as observed by $a_i$ itself) at time $t$ is represented by $s(i, t)$. Agent $a_i$'s knowledge of $a_j$'s state at time $t$ is $s'(i, j, t)$ $(i \neq j)$. The knowledge that $a_i$ has of the agents in $\mathcal{N}$ at time $t$ is represented by:

$$\mathcal{S}_i^t = \{s'(i, j, t) \mid j = 1, ..., n\}. \tag{9}$$

The corresponding set of actual state values of the agents in $\mathcal{N}$ is represented by:

$$\mathcal{S}^t = \{s(j, t) \mid j = 1, ..., n\}. \tag{10}$$

Note that in dynamic systems these states change with time. For example, the load level of a communication node varies with time. Thus, without timely updates, the known values can be different from the actual state values.

As noted earlier, agent $a_i$ uses $\mathcal{S}_i^t$ to select the subsequent agents to whom it forwards a task. This decision, in turn, affects the overall utility earned from processing tasks in the system. The exact function used by an agent to determine the subsequent agent depends on the task and the domain characteristics. In a TN, for example, an agent can select a subsequent agent for which it estimates that the average load on all nodes from that agent to the destination node is minimised. Thus, this decision has the effect of using the least congested path every time a call has to be set up which, in turn, maximises the number of calls routed in the system. Our analysis does not depend on the exact form of the decision function. Rather, it studies the delay between consecutive reinforcements of state information by a given information-sharing strategy that generate the knowledge $s'(i, j, t)$. The $s'(i, j, t)$ values act as the parameters in an agent's decision function. Hence, it can be concluded that the closer these values are to the true states (which, as stated earlier, change with time), the higher is the accuracy of the agent's decision. Intuitively, the shorter the delay in sharing information, the more up-to-date is the information maintained. Thus, the more effective an agent's decision. In the following subsections, we analyse the timeliness of information sharing by the different communication strategies.

## 5.2 Timeliness of Information Distribution in NN

In this section, we compute the delay incurred by an agent to get the state information from the others in $\mathcal{N}$ using the NN strategy. For this, we focus on $a_1$'s knowledge about the states of $\{a_1, ..., a_n\}$ at time $t$. At this time, $a_1$ has the knowledge of its actual state $s(1, t)$. However, its knowledge of the other agents $s'(1, j, t)$ $(j = 2, ..., n)$ are different from the corresponding true states by an amount equal to $\mid s(j, t) - s'(1, j, t) \mid$. Note that the information that an agent maintains at a given time is the result of the previous communication that occurred between the agents (refer to section 4.2.2 for a discussion on how a particular implementation (QR) of the NN protocol works).

Since tasks originate every $t_c$ time steps, an agent $a_j$ $(j = 1, ..., n - 1)$ requests its subsequent agent $a_{j+1}$ for the latter's knowledge every $t_c$ time steps. Following a request at any time $t$, $a_j$ receives the response from $a_{j+1}$ after a delay of $2\Delta t$, assuming the request arrives at $a_{j+1}$ after a delay of $\Delta t$, and the response of $a_{j+1}$ comes back to $a_j$ after a further delay of $\Delta t$, at $t + 2\Delta t$. Note here $\Delta t$ refers to the communication delay of a message between directly communicating agents. Referring to the discussion in section 4.2.2, we note that $a_{j+1}$ provides the information that it has of the set of agents $\{a_{j+1}, ..., a_n\}$.[9] However, $a_{j+1}$'s knowledge are based on the information it received from $a_{j+2}$ on its previous request to $a_{j+2}$. The previous request of $a_{j+1}$ to $a_{j+2}$ was during the processing of the previous task at $t + \Delta t - t_c$ for which it had received a response at $t + \Delta t - t_c + 2\Delta t$ (i.e., after a delay of $((t_c - \Delta t) - 2\Delta t)$). In a similar way, that response of $a_{j+2}$ to $a_{j+1}$ contained information that $a_{j+2}$ received from its previous request to $a_{j+3}$. That request of $a_{j+2}$ to $a_{j+3}$ was at $t + 2\Delta t - 2t_c$ for which it had received a response at $t + 2\Delta t - 2t_c + 2\Delta t$ (i.e., after a delay of $(2(t_c - \Delta t) - 2\Delta t)$). Extending this procedure to all subsequent agents, therefore, at time $t + 2\Delta t$, the information that $a_j$ has of any other subsequent agent $a_k$ is the state of $a_k$ delayed by an amount $(d - 1)(t_c - \Delta t) - 2\Delta t + \Delta t$, where $d = k - j$. Note here, an extra $\Delta t$ is added to the delay because although the response was received at $t + (d - 1)\Delta t - (d - 1)t_c + 2\Delta t$, but this contained information about $a_k$ at time $t + (d - 1)\Delta t - (d - 1)t_c + 2\Delta t - \Delta t$.

The above description is summarised in Table 1. In this table, the rows represent agents (with the agent numbers increasing from top to bottom) and the columns represent time (with time farther in the past as we move from left to right). More specifically, in Table 1, each element represents a time when an agent (represented by the row number) received the information from its subsequent agent. In particular, it focuses on agent $a_1$ and assumes that it has requested $a_2$ for the latter's knowledge at time $t$. Therefore, $a_1$ receives information from $a_2$ at $t + 2\Delta t$ (row 1, column 2 in Table 1). However, the knowledge about the subsequent agents that $a_2$ provides $a_1$ are based on the requests that these agents made at times further delayed in the past. These are the first element of each row. Hence, at time $t+2\Delta t$, the set of state information of the agents $\{a_1, ..., a_n\}$ that $a_1$ has is the following:

$$\mathcal{S}_1^{t+2\Delta t} = \{s(1, t+2\Delta t)\} \cup \{s(i, t+(i-2)\Delta t+2\Delta t-(i-2)t_c-\Delta t) \mid i = 2, ..., n\}. \quad (11)$$

---

9. Note that in section 4.2.2, we discussed QR, where a *summary* of the states of the subsequent agents is communicated. In this formalisation, we consider an agent maintains separate records of the states of other agents. Such a consideration helps explain the impact of a given communication strategy on the accuracy of an agent's knowledge.

Table 1: Time diagram for nearest-neighbour sharing

| Agent | $\leftarrow$ Time | | | |
|---|---|---|---|---|
| $a_1$ | $(t+2\Delta t)$ | $(t+2\Delta t-t_c)$ | $\cdots$ | $(t+2\Delta t-(n-2)t_c)$ |
| $a_2$ | | $(t+\Delta t+2\Delta t-t_c)$ | $\cdots$ | $(t+\Delta t+2\Delta t-(n-2)t_c)$ |
| $\vdots$ | | | | $\vdots$ |
| $a_{n-1}$ | | | | $(t+(n-2)\Delta t+2\Delta t-(n-2)t_c)$ |

Note that, an additional $\Delta t$ is subtracted in the $u$ values of all subsequent agents in (11). This is because, while Table 1 shows the last time an agent *received* information from its subsequent agent, this information is, in fact, delayed by an amount of $\Delta t$; hence, the $\Delta t$ is subtracted. For clarity, $t+2\Delta t$ in (11) is replaced by $t'$. Thus,

$$\mathcal{S}_1^{t'} = \{s(1, t')\} \cup \{s(i, t'+(i-2)\Delta t-(i-2)t_c-\Delta t) \mid i = 2, ..., n\}. \quad (12)$$

The true states of these agents, at time $t'$, are

$$\mathcal{S}^{t'} = \{s(i, t') \mid i = 1, ..., n\}. \quad (13)$$

Thus, (12) shows that the knowledge that agent $a_1$ has of any other agent in $\mathcal{N}$ is delayed by an amount that depends on the distance (number of hops) between them. More specifically, the knowledge that an agent, say $a_i$, has of another agent, say $a_j$, that is $k$ hops away is delayed by an amount:

$$t_{\texttt{delay}}^{\texttt{NN}} = (k-1)(t_c-\Delta t)+\Delta t. \quad (14)$$

This measure of delay incurred in NN will be compared to the same in PTC.

## 5.3 Timeliness of Information Distribution in PTC

In this section, we compute the delay incurred by an agent to get the state information from the others in $\mathcal{N}$ using the PTC protocol. Similar to the analysis in section 5.2, we assume that a task originates every $t_c$ time steps when $a_1$ initiates the processing of the task and forwards the request to the remaining agents $\{a_2, ..., a_n\}$. In the PTC sharing protocol, the state information of agents are communicated only after a task is completed. Since new tasks are processed every $t_c$ time steps, it can be inferred that the distribution of state information by the agents occur every $t_c$ time steps (i.e., after every task completion phase and assuming that the communication delay between any two directly communicating nodes remains the same).

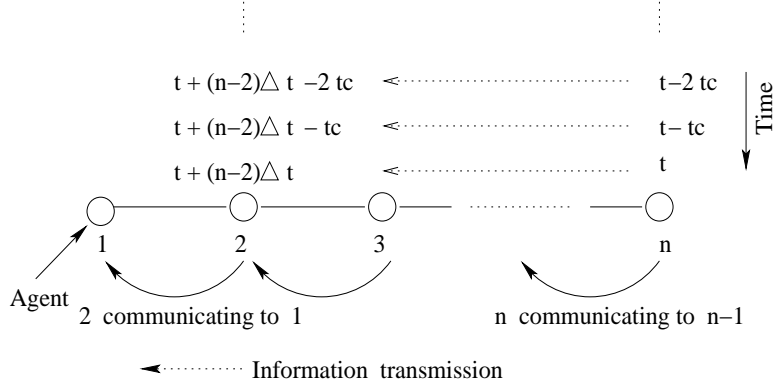

Figure 2: Time diagram for PTC sharing

Therefore, $a_n$ transmits its state information to $a_{n-1}$ at time $t$ (i.e., when a task completes at time $t$, for any value of $t$), and then at $t + t_c$, $t + 2t_c$, and so on. Given this, $a_{n-1}$ transmits its own state and the information received from $a_n$ to $a_{n-2}$ at $t + \Delta t$, and then at $t + \Delta t + t_c$, $t + \Delta t + 2t_c$, and so on (considering the delay of $\Delta t$ for the information to reach $a_{n-1}$ from $a_n$). Extending this process, it can be inferred that $a_2$ transmits its own state and its knowledge of the set of agents $\{a_3, ..., a_n\}$ to $a_1$ at $t + (n-2)\Delta t$, and then at $t + (n-2)\Delta t + t_c$, $t + (n-2)\Delta t + 2t_c$, and so on. Figure 2 shows this process. In this figure, each agent is labelled with the time at which it transmits its state information to its previous agent. Thus the state information that agent $a_1$ receives from its subsequent agent (in this case, $a_2$) contains the information of the rest of $\mathcal{N}$ delayed by multiples of $\Delta t$. Thus, $a_1$ has the following information about the subsequent agent states (assuming it received information from $a_2$ at time $t'$):

$$\mathcal{S}_1^{t'} = \{s(i, t' - (i-1)\Delta t) \mid i = 1, ..., n\}. \tag{15}$$

The true states of these agents at this time $t'$, shown in (13), however, are different from these values.

Thus, (15) shows that the information that agent $a_1$ has of any other agent in $\mathcal{N}$ is delayed by an amount that depends on the distance (number of hops) between them. More specifically, the information that an agent, say $a_i$, has of another agent, say $a_j$, that is $k$ hops away is delayed by an amount:

$$t_{\texttt{delay}}^{\texttt{PTC}} = k\Delta t. \tag{16}$$

In the following section, we use the delay measures computed in (16) and (14) to establish the advantage of PTC compared to NN.

## 5.4 Comparing Timeliness of Information Distribution in PTC and NN

The analysis of section 5.3 shows that using the PTC strategy, an agent, say $i$, after the completion of a task episode, receives the local state information of another agent $j$ after a delay of $k\Delta t$ (see formula (16)), where $k$ is the hop count between $i$ and $j$. In the NN strategy, on the other hand, agent $i$ receives $j$'s state information after a delay of $(k-1)(t_c - \Delta t) + \Delta t$ (see formula (14)). Comparing the delays, the following can be concluded.

**Proposition 1** *The delay for non-local information to reach an agent is less using PTC than using NN if the task environment periodicity is greater than the round trip communication delay of a message between directly communicating agents.*

This is because, for $k > 1$,

$$t_c > 2\Delta t \Rightarrow k\Delta t < (k-1)(t_c - \Delta t) + \Delta t. \tag{17}$$

In a typical MAS, the interval between successive episodes of task execution $(t_c)$ is much longer than the communication latency between two directly communicating agents $(\Delta t)$.[10] Hence, the delay due to PTC is, for all practical purposes, much less than that of NN.

Having established that PTC distributes information in a more time-efficient manner than NN, we now focus on analysing how this characteristic of PTC creates more up-to-date information.

## 5.5 Improved Estimation Accuracy using PTC

In this section, the improved time-efficient information distribution of PTC is mapped to the improved quality of information learned by PTC over NN. The key idea is that, the shorter the delay between successive information messages, the more accurate is the knowledge of the actual states.

As stated before, node states vary dynamically over time. However, at a given time, the state of a node can have a certain value from a certain set of values, say $\mathcal{V}$. Also, a node $n_i$ retains its state $s_m$,[11] where $s_m \in \mathcal{V}$ and $m \in \{1, \cdots, M\}$, where $M = | \mathcal{V} |$, for a certain length of time, say $l_m$. We consider that a certain node $n_i$ is dynamically estimating the states of another node $n_j$ that is at a distance of $k$ hops from $n_i$. Given the above information, we want to compute the expected value for a given number of state changes that can occur in $n_j$ in a given time duration, say $t_D$. We hypothesise that if $t_D$ increases, so does the expected value for any number of state changes within $t_D$. Therefore, if $n_i$ receives information from $n_j$ with higher delays then it loses more state-change information of the latter. Since NN has a greater delay than PTC (section 5.4), it incurs a higher loss of state-change information than PTC. In this context, we define the following:

**Definition 2** *Given a set of states $\mathcal{V} = \{s_m \mid m = 1, ..., M\}$ ($M = | \mathcal{V} |$), where each state value $s_m$ lasts for a time-length of $l_m$, and a time duration $t_D$, where $t_D < \sum l_m$, a **coverage of size $h$ from $\mathcal{V}$ on $t_D$**, represented by $c(h, \mathcal{V}, t_D)$, is a set of $h$ different states of $\mathcal{V}$ ($h \leq M$) such that $\sum\limits_h l_i \leq t_D$.*

Using the above, the expected value of $h$ different state changes of $n_j$ within a time-interval $t_D$ is given by:

$$\frac{\text{number of possible } c(h+1, \mathcal{V}, t_D)}{\binom{M}{h+1}}, \tag{18}$$

where the numerator counts all possible coverages of size $h + 1$ (hence, having $h$ different state *changes*) from $\mathcal{V}$ on $t_D$. The denominator enumerates all possible ways of choosing $h + 1$ different states from $\mathcal{V}$.

---

10. For example, in the type of communication networks we are studying in this research, the typical delay between successive calls is of the order of minutes, whereas the communication latency between adjacent nodes is of the order of milliseconds.

11. The representation $s(i, t)$ used earlier in this section to identify the state value of agent $a_i$ at time $t$ is replaced with $s_m$. In the current discussion, since we are considering one agent and the different state values that it can take, the identifier $i$ and time $t$ are dropped for an easier notation. Nevertheless, $\forall i, \forall t, \exists m, s(i, t) = s_m$.

Now, if we consider a duration $t'_D > t_D$ ($t'_D < \sum l_m$),[12] then it is trivial to identify that every $c(h, \mathcal{V}, t_D)$ will also be a $c(h, \mathcal{V}, t'_D)$, for all $h$. This is because, all combinations of $l_i$'s that "fit" within $t_D$ would necessarily fit within $t'_D$. Therefore, it can be said that $\lfloor c(h, \mathcal{V}, t'_D) \rfloor = c(h, \mathcal{V}, t_D)$. Hence, the numerator of (18) with $t_D$ replaced by $t'_D$ would at least be equal to that for $t_D$. The above reasoning brings us to the conclusion that the expected value for observing $K$ different state changes (for any $K$) increases with increasing delay between successive observations. Thus, more state-change information is lost as the delay between observations increases. Since, according to section 5.4, PTC achieves a lower delay than NN between successive observations, the following can be concluded.

**Proposition 2** *PTC incurs a lower loss of state-change information than NN.*

The analysis presented so far assumes a periodic environment where the task episodes repeat after intervals of constant length. In the following, we present a similar analysis with the periodic assumption removed and demonstrate that the same conclusions hold.

### 5.6 Non-periodic Task Environment

The analyses presented in sections 5.2 and 5.3 are based on the assumption that task completion episodes repeat every $t_c$ time steps, with a constant $t_c$. Therefore, the formulas (14) and (16) were derived using only one of these episodes. In a more general setting, however, the task processing episodes would be non-periodic, with the time between successive task completion episodes varying. In that case, these formulas have to be computed considering the successive episodes as opposed to only one. In this context, note that the information dissemination delay of NN alone (formula (14)) depends on the value of $t_c$. Therefore, the assumption of non-periodic episodes impacts the delay terms of only NN. The following discussion indicates how to account for the non-periodicity.

Considering the case described before in section 5.2 where agent $a_i$ maintains the state of $a_j$ which is $k$ hops away. In a non-periodic situation, formula (14) changes to,

$$t^{\texttt{NN}}_{\texttt{delay}} = \sum_{m=1}^{k-1} t_c^{j-m} - (k-2)\Delta t, \tag{19}$$

where, $t_c^{j-1}$ (for any $j$) represents the most recent episode, $t_c^{j-2}$ the second most recent episode, and so on.

Using a method similar to that discussed in section 5.4, $t^{\texttt{PTC}}_{\texttt{delay}}$ given by (16) can be compared to $t^{\texttt{NN}}_{\texttt{delay}}$ given by (19). Therefore, we can conclude $t^{\texttt{PTC}}_{\texttt{delay}} < t^{\texttt{NN}}_{\texttt{delay}}$ if:

$$2(k-2)\Delta t < \sum_{m=1}^{k-2} t_c^{j-m}. \tag{20}$$

The following summarises this observation.

**Proposition 3** *In a non-periodic task environment, the information dissemination delay is less for PTC than for NN if the time between any two successive task originations is greater than the round trip communication delay of a message between directly communicating agents.*

---

12. If $t_D = \sum l_m$, then the expected value of observing $h$ state-changes is equal to 1, for all $h$. If $t_D > \sum l_m$, then we can apply the same reasoning as above for the modified duration $t''_D$, where $t_D \equiv t''_D (\texttt{mod} \sum l_m)$ to reach the same conclusions.

This is because (from condition (20)), for $m \geq 1$,

$$t_c^{j-m} > 2\Delta t \Rightarrow t_{\texttt{delay}}^{\texttt{PTC}} < t_{\texttt{delay}}^{\texttt{NN}}. \tag{21}$$

Proposition 3 is similar to proposition 1. It is true for all practical purposes because the time interval between successive task processing episodes is typically much greater than the round-trip communication delay of a message between directly communicating nodes.

Since the delay between successive information received is smaller in PTC than in NN, it can be shown similar to section 5.5, that under the non-periodic task assumption, the knowledge about the non-local states generated by PTC captures the changes in these states better than that by NN.

**Proposition 4** *In a non-periodic task environment, PTC incurs a lower loss of state-change information than NN.*

The preceding analysis demonstrates that, under all practical purposes, the timeliness of information distribution and the quality of non-local state information learned by our PTC information-sharing strategy are better than the nearest-neighbour strategy under general non-periodic environments. With these theoretical results, it is reasonable to infer that PTC allows the agents to take better informed decisions than NN, which, in turn, generates better system performance. To demonstrate this further, the practical advantage of PTC is evaluated using empirical analysis in a simulated wireless telephone network. The following section describes the simulation environment.

## 6. Implementing Information-Sharing Strategies in a Simulated Wireless Telephone Network

In this section, we first enumerate a number of important physical properties and functional characteristics of this application. These properties are simulated in our system to appropriately capture their effects on its performance. Subsequently, we describe our implementation of a cooperative resource allocation system for routing calls in a circuit-switched network. In particular, it elaborates the implementations of the PTC, QR, and TPOT-RL strategies (described in section 4.2) in the simulation. Finally, we present two heuristics based on the PTC principle for aggregating state information (described in section 4.2.1).

### 6.1 Domain Properties

We assume the following characteristic properties of our simulation of a TN. These properties are typical of the broad class of wireless meshed networks (Krag & Buettrich, 2004) where a set of wireless nodes with limited communication bandwidth radio-communicate with those within their transmission range. Note that these properties correspond to the more general description of the TN domain presented in section 3.

- The communication nodes have limited bandwidth. Therefore, the number of calls that can be handled by a node is limited.

- A node can only communicate with the nodes that are within its transmission range (its immediate neighbours).

- Calls can originate/terminate at any node. These calls originate throughout the simulation and last for a finite duration (thus, indicating a continuous usage of the network).

- A node's total available bandwidth is divided into two segments: the *call channel* and the *control channel*. The former is used to route calls and the latter to communicate information and control messages.

- The resource available at a node is its call-channel bandwidth. One unit of this is allocated for each call routed via the node. The bandwidth units of the nodes on a call path are occupied throughout the duration of a call to establish a circuit. Thus, we consider a circuit-switched network where bandwidth is allocated end to end to establish calls.

- Each node is modelled as an agent. Every agent has the aim of forwarding a call to the neighbour it believes is the first node on a path to the destination with the maximum call-channel bandwidth availability.

- We define the *state* of a node at a given time as the ratio of the call channel bandwidth units it has unallocated to the maximum number of units that it can handle. Each agent has *perfect knowledge* of the state of the node it represents and *estimates* of the states of other agents.

## 6.2 Cooperative Bandwidth Allocation for Call Routing

In section 4.2, we described broadly how the routing agents allocate bandwidth in sequence to connect a call between the source and destination nodes. Here, a more detailed description of our implementation of this system is presented. To this end, we identify from the discussion of section 4.2 that the different actions of agents are in response to four types of information message: (i) request to forward a call ($m_r$), (ii) request to connect a call ($m_c$), (iii) request to drop a call ($m_d$), and (iv) request to penalise loop agents ($m_p$). In the following, these activities are elaborated. To facilitate this discussion, we first define the following set of variables used in our description: $\mathcal{N}$, set of total agents in the network $\{a_1, ..., a_N\}$, where $N = |\mathcal{N}|$; $\mathcal{K}_i$, set of $K_i$ neighbours of agent $a_i$ ($K_i = |\mathcal{K}_i|$); a Q-function $Q_i$ for each agent $a_i$, where $Q_i : \mathcal{N} \times \mathcal{K}_i \rightarrow [0, 1]$. The Q-function $Q_i(n, d)$ is $a_i$'s estimate of the bandwidth availability in the nodes on all paths to $a_d$ ($\in \mathcal{N}$) via its neighbour $a_n$ ($\in \mathcal{K}_i$); $s(i)$: actual node *state* of $a_i$; $\mathcal{B}_{j,k,...,z}$: a set of node *states* $\{s(j), s(k), ..., s(z)\}$; $E_i(d)$: feedback information provided by $a_i$ about its estimate of call channel bandwidth units available over all routes to $a_d$ from $a_i$; $E_{\mathcal{K}_i}(d)$, a set of $E_j(d)$ feedback estimates from all $a_j$ in $\mathcal{K}_i$; $R_i$, a *reward* computed by $a_i$ from node state values that it receives via communication from other nodes. In TPOT-RL, however, there is a difference in the representation of the Q-estimates from that described above. This is because, in TPOT-RL, an agent uses an action-dependent feature function, which summarises the local effects of its actions, to estimate bandwidth availability. This function is based on the bandwidth-usage level on the links connecting $a_i$ to its neighbours. Thus, if the variable $l_{i,n}$ denotes the portion of $a_i$'s total call-channel bandwidth being used for calls that are routed via its neighbour $a_n$, then $a_i$'s Q-estimate for bandwidth availability to reach a destination $a_d$ via $a_n$ is: $Q_i(e_{i,n}, n, d)$. In this, $e_{i,n}$ is "high" if the value of $l_{i,n}$, measured over a certain `activity-window` time interval in the past, is more than a certain threshold, or "low" otherwise (section 4.2.3 specifies the values used for these parameters in our experiments).

A message of type $m_r$, $m_c$, or $m_d$ contains the following information: ids of the call source ($a_s$) and destination ($a_d$), the set of ids of the nodes through which it has been routed (*path*), the time of origination ($t_o$, on the global clock) and setup ($t_{live}$, an absolute interval) of the call, and (depending on the information-sharing strategy used) a set of node state values $\mathcal{B}_{k,...,z}$. The current global time is represented by $t$. An $m_p$-type message contains, instead of the call source id, the id of the node where the loop starts. Also, on a given route, $a_{i+1}(m_T)$ ($T$ can be $r$, $c$, $d$, or $p$) returns the agent id at a position one hop closer to the call destination $a_d(m_T)$ than the current agent $a_i$, while $a_{i-1}(m_T)$ returns the id one hop closer to the call source $a_s(m_T)$. However, $a_s(m_p)$ denotes the source of the loop and not the call source node.

With this in place, Figure 3 shows the agent activities. Upon receiving an $m_r$ (line 1), an agent (say, $a_i$) checks whether it has no unallocated bandwidth or the call forwarding process has lasted beyond the maximum setup time limit (line 2). In either case, the forwarding process is stopped and $a_i$ transmits an $m_d$ (line 5), generated from $m_r$ (line 4) to refer the appropriate call represented by $m_r$, to the previous agent. Upon receiving $m_d$ (line 33), an agent frees up the pre-allocated call channel bandwidth (line 34) and sends the same $m_d$ to the previous agent (line 36) until $a_s(m_d)$ is reached.

If neither of the conditions in line 3 are satisfied, $a_i$ first checks if a loop has occurred (by checking if $path(m)$ includes its own id). If it has (line 7), $a_i$ generates an $m_p$-type message (line 8) and computes a penalty $p = (-1)0.9^{x+1}$ (line 10), where $x$ is the hop count from $a_i$ to the end of the loop (i.e., where the loop was first detected). This penalty amount is added to $m_p$ (line 11). Also, the source node id in $m_p$ is set to be the id of the node where the loop was detected (line 12). Then this $m_p$ message is transmitted to the previous agent on the loop (line 13). Upon receiving an $m_p$-type message (line 50), $a_i$ uses the penalty in $m_p$ to update its prior Q-estimate of the destination node; depending on whether the strategy used is TPOT-RL or not, one of the updates (line 52 or 54, respectively) gets executed. Subsequently, if $a_i$ is not the agent where the loop was detected (line 55), it computes a new penalty which is an exponentially decreasing function of the distance of that node from the loop end, adds this to $m_p$, and de-allocates the pre-allocated bandwidth for this call (line 57). Finally, it transmits the $m_p$ to the previous agent (line 58). The intuition here is that the further a node is from the loop end, the less it is responsible for causing the looping to occur. Hence, the lower the penalty it gets.[13]

After checking for cycles, $a_i$ then checks if it is the destination of the current call (line 14). If it is, it will allocate one bandwidth unit (line 16) and send an $m_c$ (generated from $m_r$ for reasons cited before) to the previous agent on the route of this call (line 21) to continue this process of allocation (line 42) and sending the message (line 49) until $a_s(m_c)$ is reached; at which point a complete circuit is established. However, if an agent uses the TPOT-RL strategy, before sending the $m_c$ (line 21), it stores the estimates obtained along with the $m_r$ (line 20); how such estimates are propagated in TPOT-RL is described shortly.

In the two information-sharing heuristics that we have designed based on PTC (described in sections 6.3.1 and 6.3.2), an agent $a_i$ attaches its own node state ($s(i)$) to the message $m_c$ (lines 18 and 47).[14] Each path agent $a_i$, using PTC, upon receiving an $m_c$ that con-

---

13. In our experiments, we have observed that this heuristic substantially reduces the number and size of loops. Hence, it effectively reduces wasteful use of resources, since loops represent redundant portions of a call path.

14. The PTC principle advocates information distribution only after task completion. However, a task execution process can fail (e.g., call routing failing in our example application). Note that in this situation, PTC can use the task failure as the event to trigger information distribution. We have used this concept and introduced information distribution after call failures. So far, we have observed that the BW allocation quality is better than distributing information only after call successes.

```
//for a message m in control channel Ci of agent ai
1. if(mr) // CALL FORWARD MESSAGE
2.      if (s(i) == 0|t > (to(mr) + tlive(mr))) // no available bandwidth
3.                                              // or, time exceeded setup time
4.          md ← deriveFrom(mr);//generate drop call message
5.          inform(ai−1(mr), md); // inform drop call to previous agent
6.      else
7.          if(loop(mr)) // loop detected in current call route
8.              mp ← deriveFrom(mr); //generate penalty message
9.              x ← mp.loopEndHopCount(ai);//distance of ai from the loop end
10.             penalty = (−1)0.9^{x+1};//penalty amount, x = 0 for the first loop agent
11.             mp.addPenalty(penalty);
12.             mp.setSourceNode(ai); //set the source node id to the node where the loop was detected
13.             inform(ai−1(mr), mp);//penalise previous loop agent
14.          if(ad(mr) == ai) //this node is the destination
15.              mc ← deriveFrom(mr);//generate connect call message
16.              allocateUnitBandwidth(mc);
17.              if PTC //see sections 6.3.1, and 6.3.2
18.                  append s(i) to mc;
19.              if TPOT-RL//see section 4.2.3
20.                  storeMessage(mr);// accumulate estimates
21.              inform(ai−1(mr), mc);// inform connect to previous agent
22.          else
23.              //select a neighbour based on Q-estimates for the given destination
24.              aj ← selectNeighbour(Qi, ad(mr));
25.              preAllocateCall(mr); //pre-allocate bandwidth
26.              if TPOT-RL
27.                  append Qi(ei,i+1, ad(mr), ai+1) to mr;// send own estimate
28.              inform(aj, mr);// forward call to selected agent
29.              if QR //see section 4.2.2
30.                  //selected neighbour aj returns its local estimate of bandwidth availability
31.                  Ej(ad(mr)) ← min(s(j), max_{ak∈Kj} Qj(ad(mr), ak))
32.                  Qi(ad(mr), aj) ← (1 − α)Qi(ad(mr), aj) + αEj(ad(mr));
33. else if(md) // DROP CALL MESSAGE
34.     deAllocateCall(md);//de-allocate bandwidth
35.     if(ai ≠ as(md)) //if not source
36.         inform(ai−1(md), md);
37. else if(mc) // CONNECT MESSAGE
38.     if TPOT-RL and m_c^r//reward message in TPOT-RL
39.         Ri ← computeReward(B_{ai+1}(m_c^r), ..., ad(m_c^r));
40.         Qi(ei,i+1, ad(m_c^r), ai+1(m_c^r)) ← (1 − α)Qi(ei,i+1, ad(m_c^r), ai+1(m_c^r)) + αRi;
41.     else
42.         allocateUnitBandwidth(mc);
43.         if PTC //see sections 6.3.1, and 6.3.2
44.             // compute reward using own estimate and others'
45.             Ri ← computeReward(B_{ai+1}(mc), ..., ad(mc));
46.             Qi(ad(mc), ai+1(mc)) ← (1 − α)Qi(ad(mc), ai+1(mc)) + αRi;
47.             append s(i) to mc;
48.     if(ai ≠ as(mc)) //if not source
49.         inform(ai−1(mc), mc);
50. else if(mp) // PENALTY MESSAGE
51.     if TPOT-RL
52.         Qi(ei,i+1, ad(mp), ai+1(mp)) ← (1 − α)Qi(ei,i+1, ad(mp), ai+1(mp)) + αmp.penalty;
53.     else
54.         Qi(ad(mp), ai+1(mp)) ← (1 − α)Qi(ad(mp), ai+1(mp)) + αmp.penalty;
55.     if(ai ≠ as(mp))
56.         d ← mp.loopEndHopCount(ai);//distance of ai from the loop end
57.         penalty = (−1)0.9^{d+1}; mp.addPenalty(penalty); deAllocate(mp);//de-allocate bandwidth
58.         inform(ai−1(mp), mp);
59. if TPOT-RL //tasks specific to TPOT-RL
60.     monitorLinkUsage(Ki); //measure usage of all li,j, aj ∈ Ki
61.     if(t % update-interval == 0)
62.         for all accumulated mr
63.             m_r^r ← deriveFrom(mr); //create reward message
64.             inform(ai−1(mr), m_c^r); //transmit upstream along this mr's route
65. t ← t + 1;
```

Figure 3: Agent actions in response to various message types

tains the node state information transmitted from other agents, computes a reward $R_i$ as a function of the set of communicated state values $\mathcal{B}_{a_{i+1}(m_c),...,a_d(m_c)}$ contained in $m_c$ (line 45). Sections 6.3.1 and 6.3.2 define two heuristics for calculating this reward value. In either case, however, the reward is used to update the prior Q-estimates (line 46). The update rule follows standard Q-learning, where $\alpha$ is the learning rate (see section 4.1). Thus, the agents cooperatively volunteer their local information to one another to improve their estimates of the unobserved node states.

On the other hand, if $a_i$ is not the destination for this call, it selects one of its neighbours (excluding the one from which it received the $m_r$) to forward the call request (line 24). This is done by defining a probability distribution over $a_i$'s set of Q-estimates of its neighbours. In particular, the probability of selecting a neighbour $a_j$ is given by:

$$Pr(a_j) = \frac{\exp(\frac{Q_i(a_d(m),a_j)}{\tau})}{\sum\limits_{a_k \in \mathcal{K}_i, a_k \neq a_{i-1}(m_r)} \exp(\frac{Q_i(a_d(m),a_k)}{\tau})}. \tag{22}$$

Note that equation 22 refers to the selection mechanism when PTC or QR is used. If TPOT-RL is used, then, $Q_i(e_{i,j}, a_d(m), a_j)$ replaces the Q-values in equation 22. Here $\tau$ is the "temperature" parameter (range: [0,1]) and controls how much the relative differences between various Q-estimates would affect the relative probabilities of selection (the smaller the $\tau$, the larger the skewness). This is a standard heuristic called Boltzmann exploration (Watkins, 1989) to probabilistically choose between alternative options. Subsequently, $a_i$ pre-allocates a bandwidth unit of its call channel (line 25) and forwards $m_r$ to $a_j$ (line 28). In the QR strategy, the selected neighbour $a_j$ responds to $a_i$ with its own estimate $E_j(a_d(m_r))$ of bandwidth availability on routes to the destination $a_d(m_r)$ (line 31) which is equal to $\min(s(j), \max\limits_{a_k \in \mathcal{K}_j} Q_j(a_d(m_r), a_k))$ (refer to section 4.2.2). The requesting agent $a_i$ uses this estimate $E_j(a_d(m))$ to update its prior estimate $Q_i(a_d(m), a_j)$ (line 32). In TPOT-RL, $a_i$ appends its Q-estimate to the $m_r$ (line 27) before forwarding the latter to $a_j$. This is how the sequence of estimates gets propagated along with the call forwarding request in TPOT-RL.

Now we describe a number of activities that an agent performs only if TPOT-RL is used. First, $a_i$ monitors the link usage levels for all its neighbours (line 60). This information is used to compute the value of $e_{i,j}$ ($j \in \mathcal{K}_i$), as described before. Second, every `update-interval` time steps (line 61), $a_i$ starts sending reward messages along the paths of those calls that terminated at $a_i$ during that period (for those $m_r$'s that it had stored during that period). Specifically, these reward messages are analogous to the $m_c$ type message, except that no bandwidth is allocated when an agent receives one (as opposed to bandwidth being allocated when an $m_c$ is received (line 38)). To distinguish from $m_c$, we denote these messages as $m_c^r$ for our description in Figure 3. Every `update-interval` stored over the past interval, $a_i$ creates an $m_c^r$ (line 63) and sends this to the neighbour (line 64) which is the immediate upstream node along that call path. Upon receiving an $m_c^r$, an agent computes a reward using an aggregation of the information of the subsequent path nodes (line 39) (similar to PTC) and updates its Q-values with this reward (line 40) before sending the $m_c^r$ upstream. It should be noted here, the information used by the agents to compute the reward in TPOT-RL are the Q-estimates that the agents had appended while forwarding the $m_r$. Therefore, the set $\mathcal{B}_{a_{i+1}(m_c^r),...,a_d(m_c^r)}$ in line 57 represents *estimates* and not actual node states. We have used the same aggregation method for TPOT-RL as described in section 6.3.2.

The above discussion corresponds to a more detailed description of the general domain description of section 3. As identified in that description and observed in the above system description, the decision to select a specific neighbour to forward a call is critical in determining how effective the system is in successfully routing calls. Since this decision is taken based on the Q-estimates, the more accurately they reflect the true node bandwidth availabilities, the better informed are the decisions taken by an agent. It is emphasised that the information-sharing strategy plays a key role in determining the estimation accuracy. In the following, we formulate two simple heuristics of PTC that are used to define the *computeReward* function in our simulations.

## 6.3 PTC Sharing Heuristics

The discussion on agent interactions in section 6.2 explains how the agents using PTC delay transmitting the information until a call is connected. In this case, the agents along the call path can aggregate the information received from those "downstream" and pass on that information to the previous path agent. We have formulated two simple heuristics for information aggregation based on PTC, viz., to *average* the state estimates (termed PTC-A) and to take the *minimum* state estimate (termed PTC-M). These are described in the following.[15]

### 6.3.1 PTC INFORM AVERAGE CAPACITY (PTC-A)

In Figure 3, upon receiving the $m_c$, an agent using the PTC-A heuristic computes a reward value $R_i$ by *averaging* the states of all nodes on the route from $a_i$ to $a_d$ as:

$$R_i = \frac{\sum\limits_{k \in \{a_{i+1}, ..., a_d\}} s(k)}{L(a_i, a_d)}, \tag{23}$$

where $L(a_i, a_d)$ is the hop count on this route from $a_i$ to $a_d$. This describes how the function *computeReward* of Figure 3 is implemented. Subsequently, this reward is used to update its prior Q-estimate. Thus the estimates are updated with the information about the resource usage on the "downstream" nodes on the path of this call.

### 6.3.2 PTC INFORM MINIMUM CAPACITY (PTC-M)

This is similar to PTC-A: but instead of average available capacity, the *minimum* available capacity is used as reward. For example, agent $a_i$ using the PTC-M heuristic computes the reward as:

$$R_i = \min(s(i+1), ..., s(d)). \tag{24}$$

This is a more conservative estimate of bandwidth availability than the average capacity model. Thus it has the advantage that the probability of a dropped call due to agents overestimating the bandwidth availability is reduced. This heuristic is also used to aggregate estimates in TPOT-RL in our experiments.

---

15. Here, it should be noted that the formal analysis in section 5 is based on the agents having separate estimates of individual agents. Maintaining an aggregate estimate on a set of agents, however, reduces the computational complexity at decision time in our simulations. Alternatively, for example, with estimates of individual nodes, the run-time complexity of determining the least cost path between any two nodes is quadratic in the number of nodes in the graph (the worst case run time of Dijkstra's single-source shortest path algorithm Cormen, Leiserson, Rivest, & Stein, 2001).

## 7. Experimental Evaluation

Based on the scenario discussed in section 6, we have conducted a series of experiments to empirically evaluate the effectiveness of PTC compared to the benchmarks (QR and TPOT-RL). In this section we start by enumerating the measures chosen as indicators of system performance. Subsequently, the results obtained from the experiments on these measures are analysed.

### 7.1 Performance Measures

The following measures are chosen to evaluate the performance of a given communication strategy in improving learning in a TN.

#### 7.1.1 Number of Successful Calls

The overall objective of the cooperative agent network is to maximise the number of successfully routed calls, given the set of resources (call-channel bandwidth) available, the rate at which new calls originate, and the duration calls remain connected (hold up resource). Thus, the average number of successful calls determine how successful a learning algorithm has been given the above parameter values. In our system, we record the total number of calls that have originated ($NO$) and the total number of those calls that have been successfully routed ($NC$). Thus, the value $x = \frac{NC}{NO}$ determines the fraction of successful calls routed over the time period in which the measurements are taken.

In addition to measuring the number of calls successfully routed, we keep track of the number of calls that could be connected if the agents had global knowledge about the network bandwidth availability and if call routing could be done instantaneously. This is computed by globally searching for the availability of a path for each call. This search is done instantaneously at the beginning of each simulation time step. We term this essentially idealistic procedure the "**I**nstantaneous **Z**ero **D**elay **S**earch" (IZDS). Since, in practice, it takes a finite number of time steps to connect a call, the IZDS is repeated at every time step until either it finds a path or the call is dropped/connects. [16] In case of the former outcome, the $NC_{izds}$ count is incremented by 1. The value $x_{izds} = \frac{NC_{izds}}{NO}$ gives a hard upper limit to the call success rate under the given conditions.[17] The learning strategies are compared both against the absolute success rate ($x$) values and the percentage deviation of success rate from IZDS, $\frac{x_{izds} - x}{x_{izds}}$. It is infeasible for any practical system to attain the IZDS success rate because, in practice, there is a finite amount of delay to connect a call as opposed to the instantaneous connection in IZDS. Also, the agents in a practical system attempt to connect a call by forwarding it one hop at a time based on their individual estimates of the world states. IZDS, on the other hand, takes a global and accurate view of the entire network to find the least cost path.

---

16. Thus, IZDS is guaranteed to find a path if one is found by the actual routing algorithm, but not necessarily vice versa.

17. Note, IZDS is not an optimal measure. Since it tries to connect a call as soon as one originates, it is essentially a greedy strategy. Some other scheduling strategy, say that uses some form of lookahead before attempting to place a call, may outperform IZDS.

### 7.1.2 Successful Routes of Different Lengths

The call success rate metric, described in section 7.1.1, measures the overall rate of successful calls in the system. This is an important performance measure because it indicates how the system performs at a broad scale. So it is an indicator of how successful a communication strategy is in improving the system performance. However, it does not indicate how effective the strategy is at connecting calls at a given distance (since the success rate metric counts all calls in the system).

In our case, calls can be required to be routed to destination nodes that are at various distances from the nodes of origin. Calls destined for nodes that are at short distances are relatively easier to route than those further away. This is because more accurate estimates of the load at nodes that are nearer can be maintained. As shown in the analysis of section 5, the farther an agent, the longer is the delay for the information to arrive and, thus, the less up-to-date are the estimates. Therefore, it is less probable for a call to be routed successfully to such a distant node. A communication strategy that allows better success rates at longer distances can, therefore, be considered more competent than another that achieves a poorer success rate at long distances (all other conditions being equal).

Thus we measure the number of successful call connections for various distances between the call source and destination nodes. More specifically, the minimum hop count (say, $d$) between the source and destination of a call is computed (global knowledge of the network topology is used to measure this distance) and the success counter ($NC_d$) of calls at distance $d$ is incremented if such a call is successfully connected. In this manner, a success rate for calls at a distance $d$ can be computed as $x_d = \frac{NC_d}{NO_d}$, where $NO_d$ stands for the number of calls originated with a source-destination distance of $d$. For different values of $d$, therefore, the different $x_d$ values could be used to compare the ability of different communication strategies to place calls at different lengths.

### 7.1.3 Reward Information Messages

The agents in our system use different message types for communication (section 6.2 enumerates these). However, the most important among these are the ones that carry the reward information by which the local knowledge of one agent is transmitted to another. This is because using this information, the agents update their prior estimates of the network load level. Therefore, these messages contribute directly to the quality of learning and to the overall performance of the system in allocating resources to place calls. In QR, for example, this is the message that an agent receives from its neighbour after handing over to the latter a call forwarding request (see section 4.2.2 for details). Therefore, at each call forwarding step, a new message is generated by the contacted agent and transmitted to the contacting agent. In the PTC-based models, on the other hand, these are the $m_c$ type messages, transmitted after a call connects, that contain the summary reward information (sections 6.3.1 and 6.3.2 have the details). Note, this is only one message generated by the destination node after every successful routing and transmitted upstream along the call path. [18] Although several other types of messages (e.g., $m_r$, $m_p$, and $m_d$) are exchanged between the agents, it is the messages that contain the reward that affect the learning quality the most. On the other hand, in our implementation of TPOT-RL, an agent, while forwarding a call, transmits along with the call, its own estimate of the bandwidth availability along paths to the call destination. Subsequently, this information is used to update the Q-values of the agents (see Section 7.2 for details). Therefore, these messages affect the learning of agents in the way that $m_c$ does in PTC. Hence, the message rate for TPOT-RL is measured by counting these messages.

---

18. We have recently conducted a study where PTC shares information also after a call fails to connect. This modification, obviously, increases the message rate although still keeping it lower than that of QR and TPOT-RL. However, it allows better bandwidth allocation than the PTC reported in this article and also deals with failure detection, something the current PTC is not capable of (Dutta et al., 2005).

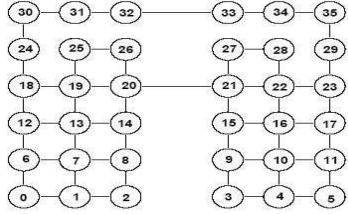

Figure 4: 36 node irregular grid

The number of such messages can, therefore, be used as a measure of the efficiency of a given communication strategy — the lower the number of messages, the higher is the efficiency (assuming a given value of some other performance measure such as call success rate). More specifically, the total number of information exchanges (represented by $m$, say) for transmitting the agents' local knowledge to one another is computed at every $T$ time steps during a simulation. The value $r = \frac{m}{T}$, therefore, gives the rate of messages transmitted in the entire system during the interval $T$. The total simulation is divided into several intervals and the values of $r$ over each such interval generate an overall time-variation of the message rate. The different communication strategies are then compared against various message rates.

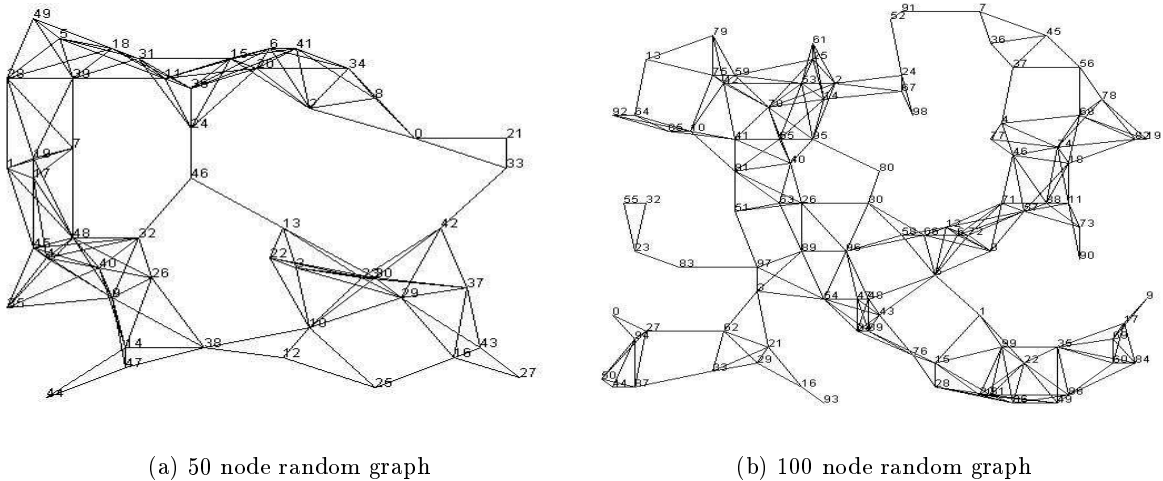

(a) 50 node random graph          (b) 100 node random graph

Figure 5: Random network topologies

## 7.2 Results and Analysis

In this section, the performance of QR, PTC-A, PTC-M, and TPOT-RL are compared against the measures described in section 7.1.

Experiments are conducted on a number of different network topologies. In the following, we report our results and observations based on some of those. Figures 4 and 5 show, respectively, a 36-node irregular grid and two randomly generated topologies — a 50-node random graph (Figure 5(a)), and a 100-node random graph (Figure 5(b)). The 36-node irregular grid topology has been used in previous papers on the application of RL in network routing (see section 2.2 for a discussion on these papers) and thus we choose it to make valid comparisons. To verify our conclusions across a wider range of topologies, we tested the same against random graphs (figures 5(a) and 5(b) show two such examples) of different sizes.[19] In all figures, the nodes are numbered for ease of reference. The edges between nodes indicate that those nodes are within each other's radio range. Note that the random graphs are designed such that any node is linked to only those within a certain maximum radial distance which simulates the transmission range of wireless nodes.

The following parameter values were used for all experiments reported henceforth (unless otherwise stated): learning rate $\alpha = 0.03$, Boltzmann exploration temperature $\tau = 0.1$, call setup time of 36, 50, and 100 time steps for the topologies in figures 4, 5(a) and 5(b), respectively. We allowed for a larger call setup time in the bigger topologies to give allowance for the larger size of the networks. An average call duration of 20 times the setup time was used. The "load" in the network is set by assigning a probability with which calls originate at every time step in the network. This call origination probability was varied to study the effect of different network loads on the performance. Calls were allowed to originate and terminate on any randomly selected node. We have tested the strategies using both (i) a constant-load simulation where the call origination probability is maintained the same throughout a simulation run, and (ii) a dynamically changing load simulation where the call origination probability is changed during the course of a simulation run. We have used different numbers and values of the call origination probability changes in a single simulation run to test the effect of various degrees of load fluctuations on the performance of PTC, QR, and TPOT-RL. In these dynamic settings, the call origination probability is changed at equal intervals in a simulation run. A single simulation run lasted for 500,000 time steps for the topology in Figure 4, for 1,000,000 time steps for the topology in Figure 5(a), and for 2,000,000 time steps for the topology in Figure 5(b). Results are averaged over 10 simulation runs (these figures are statistically significant at the 95% confidence level). Also, every node had a maximum call channel capacity of 10 units.

The following sections discuss the results on the overall success of the various communication heuristics in connecting calls (in section 7.2.1), the effectiveness of these heuristics in successfully connecting long-distance calls (in section 7.2.2), and the overhead due to communicating messages (in section 7.2.3).

### 7.2.1 PERFORMANCE — CALL SUCCESS RATE

We anticipate that the higher the accuracy of learned estimates of unobserved states, the more capable the agents will be of routing calls to the destination via the most appropriate paths. Hence, in turn, the higher will be the call success rate and the lower the deviation from the IZDS success rate. In the following, the results from constant load are presented first followed by those obtained from dynamically varying load.

---

19. Various other topologies were used with varying number of nodes and connectivity patterns and the same general trends in the results were observed. Hence, here, we report on three sample topologies.

**Constant Load.** We experimented with all strategies to calculate: (i) the call success rates, and (ii) the percentage deviation of the measured success rate from the IZDS, under steady state conditions (i.e, when the call throughput in the system reached a steady value).[20] The average success rate and the average IZDS success rate, both computed during the steady state phase of the simulation, are further averaged over 10 simulation runs for a given value for the call origination probability. Further, call success rate is measured against various call origination probabilities to test the impact of network load on the success rate. These measurements are repeated for each of the three topologies.

Table 2: Call success rates for all strategies — topology of Figure 4

| Load | QR | | | PTC-A | | | PTC-M | | | TPOT-RL | | |
|---|---|---|---|---|---|---|---|---|---|---|---|---|
| | Avg | Stdev | IZDS | Avg | Stdev | IZDS | Avg | Stdev | IZDS | Avg | Stdev | IZDS |
| 0.1 | 50.94 | 0.0048 | 72.66 | 50.98 | 0.0076 | 72.11 | **51.85** | 0.0052 | 70.25 | 25.74 | 0.0022 | 99.57 |
| 0.2 | 32.41 | 0.0034 | 60.69 | 32.64 | 0.0037 | 60.56 | **33.02** | 0.0039 | 58.86 | 19.94 | 0.0016 | 97.65 |
| 0.4 | 19.99 | 0.0018 | 55.37 | 20.27 | 0.0021 | 55.23 | **20.38** | 0.002 | 52.79 | 14.82 | 0.0009 | 89.44 |
| 0.6 | 14.87 | 0.0012 | 54.87 | 15.07 | 0.0011 | 53.44 | **15.08** | 0.0014 | 50.6 | 11.52 | 0.0012 | 85.2 |

Table 3: Call success rates for all strategies — topology of Figure 5(a)

| Load | QR | | | PTC-A | | | PTC-M | | | TPOT-RL | | |
|---|---|---|---|---|---|---|---|---|---|---|---|---|
| | Avg | Stdev | IZDS | Avg | Stdev | IZDS | Avg | Stdev | IZDS | Avg | Stdev | IZDS |
| 0.1 | 57.31 | 0.0042 | 81.75 | 58.15 | 0.0052 | 81.67 | **58.49** | 0.0052 | 80.9 | 12.18 | 0.0027 | 99.99 |
| 0.2 | 37.11 | 0.002 | 69.75 | 37.32 | 0.0028 | 68.73 | **37.65** | 0.0026 | 68.33 | 10.21 | 0.0011 | 99.8 |
| 0.4 | 22.98 | 0.0012 | 61.44 | 23.35 | 0.0012 | 62.6 | **23.51** | 0.0013 | 60.43 | 7.46 | 0.0007 | 99.8 |
| 0.6 | 17.17 | 0.001 | 58.8 | 17.47 | 0.0009 | 61.29 | **17.49** | 0.0008 | 57.53 | 6.11 | 0.0004 | 99.8 |

Table 4: Call success rates for all strategies — topology of Figure 5(b)

| Load | QR | | | PTC-A | | | PTC-M | | | TPOT-RL | | |
|---|---|---|---|---|---|---|---|---|---|---|---|---|
| | Avg | Stdev | IZDS | Avg | Stdev | IZDS | Avg | Stdev | IZDS | Avg | Stdev | IZDS |
| 0.1 | 35.79 | 0.0032 | 66.52 | 35.86 | 0.0034 | 65.58 | **37.34** | 0.003 | 62.65 | 8.53 | 0.0031 | 99.88 |
| 0.2 | 24.17 | 0.0017 | 57.41 | 24.56 | 0.002 | 59.18 | **24.8** | 0.0014 | 54.27 | 6.12 | 0.0013 | 99.83 |
| 0.4 | 15.88 | 0.0011 | 54.01 | 16.16 | 0.0007 | 55.53 | **16.18** | 0.0008 | 50.72 | 4.79 | 0.0006 | 99.9 |

In more detail, Table 2 shows the average steady state success rates achieved by the different strategies and by IZDS used alongside these strategies for different network loads

---

20. Note, our TN application represents a dynamic system where node bandwidth availability changes with time (new calls are placed and existing calls terminate). The Q-values estimate the bandwidth availability. So, as bandwidth availability changes, so do the Q-values. However, the call success rate is an overall system measure which reaches a steady state when a constant call origination probability is used.

(call origination probabilities) in the topology of Figure 4. Tables 3 and 4 show the same measurements for the topologies in Figure 5(a) and 5(b), respectively. The results indicate that the average steady state call success rate (shown under column "Avg" in these tables) achieved with PTC-M dominates that of the other strategies under different network loads. For example, in Table 2, when the load is 0.1, PTC-M achieves a steady state average call success rate of 51.85%, PTC-A achieves 50.98%, QR 50.94%, and TPOT-RL 25.74%. As indicated by the analysis of section 5, PTC maintains more up-to-date information of the network states. Thus, using PTC, the agents are capable of taking better informed decisions of forwarding a call which, in turn, ensures a higher likelihood of successful connections. This is reflected in the (statistically significant) higher call success rate achieved by PTC-M over all other strategies. In particular, since the minimum-capacity heuristic restricts overestimation of the node bandwidth availability, it generates a slightly better success rate than the average capacity heuristic.

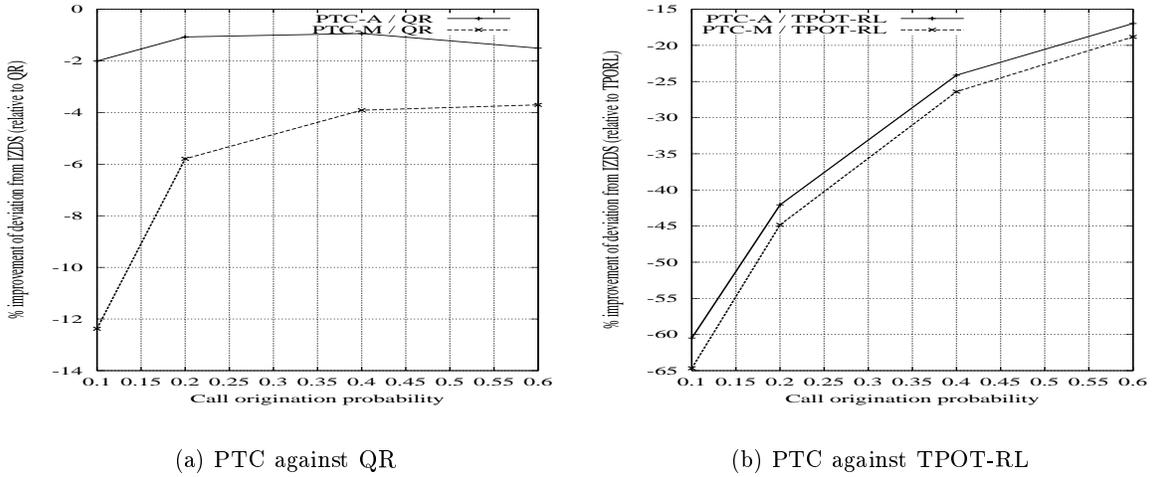

(a) PTC against QR         (b) PTC against TPOT-RL

Figure 6: Improvement of call success rate deviation from IZDS relative to QR and TPOT-RL — topology of Figure 4

We have also measured the improvements in the success rate values achieved from the PTC-based information strategies relative to those of QR and TPOT-RL. To do so, first the success rate deviation from IZDS is computed as $p = \frac{x_{izds} - x}{x_{izds}}$ (see section 7.1.1). Subsequently, $p_{PTC-M}$ and $p_{PTC-A}$ are compared to $p_{QR}$ and $p_{TPOT-RL}$ as: $\frac{p_{QR} - p_{PTC}}{p_{QR}}$ and $\frac{p_{TPOT-RL} - p_{PTC}}{p_{TPOT-RL}}$, respectively. The graphs in Figure 6 show the relative improvements of the average success rate deviation from IZDS achieved by using PTC over QR (Figure 6(a)) and over TPOT-RL (Figure 6(b)) when the topology in Figure 4 is used. Note that in this figure *a greater negative value indicates less of a deviation from the IZDS relative to QR or TPOT-RL, and, hence, a better performance of PTC.* From these graphs, a significant improvement is observed in PTC-M relative to both QR and TPOT-RL under different values of network load (statistical significance is tested at the 95% confidence level). For example, (see the row corresponding to "Load" = 0.1 in Table 2) with a call origination probability of 0.1, the deviation of the call success rate from the IZDS is 26.2% (= $\frac{70.25 - 51.85}{70.25}$) for PTC-M and 29.9% (= $\frac{72.66 - 50.94}{72.66}$) for QR. Therefore, the success rate due to PTC-M is 12.37% ($=|\frac{26.2-29.9}{29.9}|$) closer to IZDS relative to that of QR. The success rate deviation from the IZDS for PTC-A also remains lower relative to QR. Similarly, comparing PTC-M with TPOT-RL at the same load level, PTC-M achieves a 26.2% deviation from IZDS while the corresponding figure for TPOT-RL is 74.15% ($=\frac{99.57-25.74}{99.57}$). Therefore, the success rate due to PTC-M is 64.66% ($=|\frac{26.2-74.15}{74.15}|$) closer to IZDS relative to that of TPOT-RL. The performance of PTC-M dominates that of PTC-A (although both perform better than QR and TPOT-RL) because of the conservative nature of the minimum capacity heuristic compared to the average capacity (as stated above). These observations further strengthen our analysis in section 5 that by providing better quality estimates, PTC performs closer to the IZDS than QR. An additional observation is that with increasing load, the relative improvement of PTC over QR and TPOT-RL reduces. Thus, in Figure 6(a), the improvement of PTC-M over QR is 12.37% with load of 0.1, while it is only 3.7% with load of 0.6. Similarly, in Figure 6(b), PTC-M is 64.66% better than TPOT-RL at load 0.1, but 18.82% better at load 0.6. Note that an increase in load implies reduction in the time between successive task processing episodes since calls originate more frequently with increased load. As explained in section 5, the smaller the value of the interval between successive task processing episodes, the less up-to-date are the estimates generated. Therefore, with an increase in the load, the performance differences between PTC and the other strategies decrease.

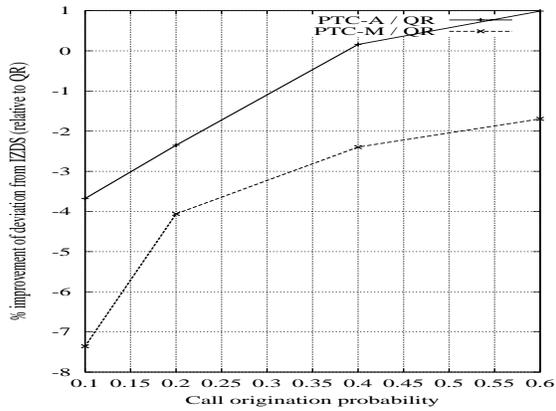

(a) PTC against QR

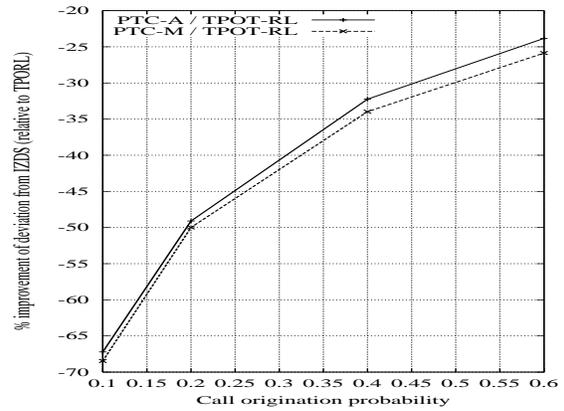

(b) PTC against TPOT-RL

Figure 7: Improvement of call success rate deviation from IZDS relative to QR and TPOT-RL — topology of Figure 5(a)

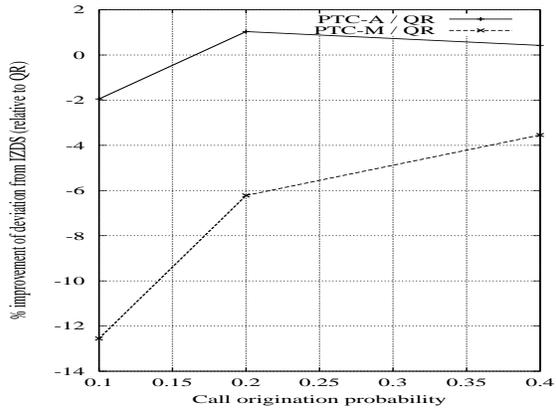

(a) PTC against QR

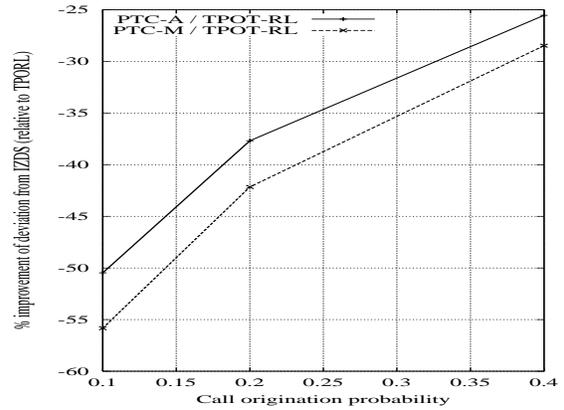

(b) PTC against TPOT-RL

Figure 8: Improvement of call success rate deviation from IZDS relative to QR and TPOT-RL — topology of Figure 5(b)

We observe identical trends in the performances of PTC-M relative to the benchmarks with the other topologies. Figure 7 shows the results for the topology in Figure 5(a), and Figure 8 for the topology in Figure 5(b).

**Dynamically Changing Load.** In contrast to the steady-state call success rate measured with constant load, here we present the time-variation of the call success rate as the network load fluctuates. This captures the responsiveness of the system call success rate to dynamically changing load levels given a particular information-sharing strategy. To this end, Figure 9(a) shows the time-variation of the call success rates of PTC-M, QR, and TPOT-RL as the call origination probability is increased from 0.1 to 0.6 in a simulation run using the topology of Figure 4.[21] It demonstrates that the PTC (in this case we show PTC-M; PTC-A is excluded from the results since it performs slightly worse than PTC-M and better than QR and TPOT-RL) call success rate remains the highest both when the load level is 0.1 (before time = 50) and when it increases to 0.6 (after time = 50). The call success rates of all strategies suffer a drop with the increase in load level since the nodes have only limited bandwidth to allocate calls, a result that we have observed previously with constant load (see Table 2). We further experimented with multiple load fluctuations over a simulation run. Figure 9(b) shows the results when the load level changes between five different values: 0.1, 0.2, 0.4, 0.6, and 0.8 in that order. In all cases, PTC is observed to have the highest call success rate out of PTC-M, QR, and TPOT-RL. Note that in these results, we have used two different types of load fluctuations: one large increase (0.1 to 0.6) and monotonically increasing. There can of course be several other patterns of load variations, such as random fluctuations or following specific probability distributions (e.g., Poisson) but these are not considered in this work. In this context, we identify that the impact of a load change is that the agents have to re-learn the new environmental condition so that calls can be placed effectively. The most severe case in this situation is that of increase in load because a higher load demands more efficient allocation of (limited) resources. Thus, having shown that PTC outperforms the benchmarks under this condition, we envisage that similar broad trends would be observed for other patterns.

---

21. We have experimented with all the other topologies under conditions of dynamically changing load. The broad patterns observed in the results are identical across all of them. Hence, we choose one sample topology to report our results in this paper.

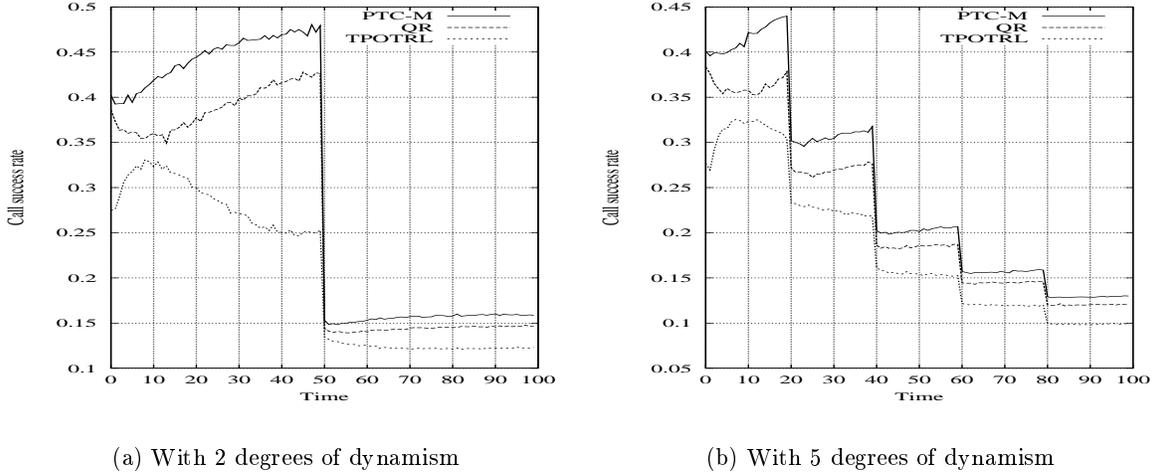

(a) With 2 degrees of dynamism          (b) With 5 degrees of dynamism

Figure 9: Time variation of call success rates of QR, PTC-M, and TPOT-RL with network load fluctuations — topology of Figure 4

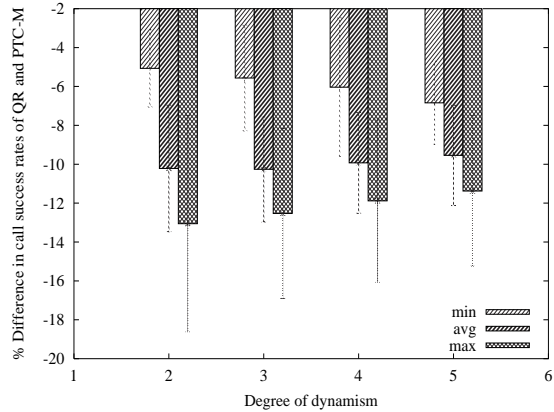

Figure 10: Summary statistics of call success differences between QR and PTC-M with network load fluctuations — topology of Figure 4

Now we aim to summarise the effects of dynamically changing load on the network call success rate given an information-sharing strategy. In so doing, we first designate the number of load levels in a simulation run as the "degree of dynamism". For example, Figure 9(a) has a degree of dynamism of 2 and Figure 9(b) has 5 degrees of dynamism. Then, for a given degree of dynamism, we compute the percentage difference of the call success rates of QR or TPOT-RL using PTC-M as the baseline as: $\frac{r_{QR}^t - r_{PTCM}^t}{r_{PTCM}^t}$, and $\frac{r_{TPOTRL}^t - r_{PTCM}^t}{r_{PTCM}^t}$, where $r^t$ is the time-varying call success rate. Note that this difference measure is also time-dependent. To summarise the improvement of call success rate using PTC-M, we find the minimum, the mean, and the maximum of this difference over every time interval during which the load-level remains constant. These statistics present the call success rate improvement range within the time interval when the load remains at a certain level. Subsequently, the means and the standard deviations of each of the minimum, the mean, and the maximum differences over all such intervals are computed. This step generates the summary of the different improvement ranges across all such intervals. For example, in Figure 9(a), the minimum, mean, and maximum percentage differences are computed in the interval where load = 0.1 and in the interval where load = 0.6. Subsequently, the mean and standard deviations of these two sets of difference statistics are computed. Figure 10 shows the above-mentioned measures for all degrees of dynamism used in our experiments when QR is compared to PTC-M. Since TPOT-RL is always observed to achieve a lower call success rate rate than both PTC-M and QR, we exclude the summary comparison of TPOT-RL with PTC-M in this paper. Along the horizontal axis of Figure 10 is the degree of dynamism while the summary statistics of the percentage success rate difference are along the vertical axis. In this figure, a degree of dynamism 2 indicates the load level is changed from 0.1 to 0.2; 3 indicates a change of 0.1, 0.2, 0.4; 4 indicates 0.1, 0.2, 0.4, 0.6; and 5 indicates 0.1, 0.2, 0.4, 0.6, 0.8. The negative values of the call success rate differences indicate $r_{QR}^t$ is lower than that of $r_{PTCM}^t$. The figure shows that the gap between the minimum and the maximum differences reduces with the number of degrees of dynamism. For instance, this gap is approximately 8% with degree of dynamism 2, and it is about 4.5% when the degree of dynamism is 5. This is because, with a higher number of load changes within a given simulation run, there is less time for any strategy to re-learn the changes in the environment. Hence, the call success rates do not attain their steady state values (as in Table 2). Nevertheless, the average statistics of the call success rate differences show that PTC-M achieves a significantly higher call success rate across all degrees of dynamism. For example, this is about 10.2% with 2 degrees of dynamism and 9.5% with 5 degrees of dynamism.

These observations reinforce our hypotheses that, under the given simulation environment, *the post-task-completion information sharing model with minimum-capacity reward achieves the best call success rate among all strategies under different network loads under conditions of both static and dynamically fluctuating loads.* For example, with constant load, the deviation of the success rate from the IZDS is up to 12.37% lower in PTC-M than QR, and up to 65% lower in PTC-M than TPOT-RL. Further, with dynamically changing load, PTC-M achieves an average 10% improvement in call success rate over QR with five different load-level changes in a simulation run.

### 7.2.2 Performance — Success Rate for Calls of Different Lengths

The call success rate values reported in section 7.2.1 indicate better performance of PTC-M over QR and TPOT-RL. In addition, to measure the effectiveness of an information-sharing heuristic in connecting a call to a destination that is at a given distance from the source, the measure $x_d$ (see section 7.1.2 for its definition) is measured for increasing values of $d$. Results from the constant load experiments are reported in this section. With dynamically changing load, the summary statistics (as described in section 7.2.1) of call success rates indicated better performance of PTC than the benchmarks for all values of $d$. Since we have already shown better success rate of PTC with dynamic load condition in section 7.2.1, we have excluded the call success rates at different distances under the same conditions in this section.

Table 5: Call success rates at various distances — topology of Figure 4

| Load | Strategy | Min hop count | | | | | | | | | |
|---|---|---|---|---|---|---|---|---|---|---|---|
| | | 1 | 2 | 3 | 4 | 5 | 6 | 7 | 8 | 9 | 10 |
| 0.1 | QR | **0.87** | **0.76** | 0.63 | 0.51 | 0.42 | 0.37 | 0.355 | 0.343 | 0.344 | 0.34 |
| | PTC-A | 0.863 | 0.748 | 0.632 | 0.516 | 0.426 | 0.374 | 0.355 | 0.345 | 0.348 | 0.354 |
| | PTC-M | 0.857 | 0.753 | **0.643** | **0.534** | **0.438** | **0.381** | **0.362** | **0.356** | **0.353** | **0.365** |
| | TPOT-RL | 0.633 | 0.532 | 0.356 | 0.218 | 0.164 | 0.148 | 0.126 | 0.094 | 0.071 | 0.059 |
| 0.2 | QR | **0.785** | **0.592** | 0.44 | 0.31 | 0.222 | 0.178 | 0.158 | 0.147 | 0.147 | 0.139 |
| | PTC-A | 0.773 | 0.583 | 0.44 | 0.315 | 0.231 | 0.181 | 0.164 | 0.155 | 0.151 | 0.151 |
| | PTC-M | 0.768 | 0.583 | **0.444** | **0.326** | **0.237** | **0.187** | **0.168** | **0.155** | **0.151** | **0.152** |
| | TPOT-RL | 0.584 | 0.462 | 0.266 | 0.139 | 0.101 | 0.098 | 0.086 | 0.066 | 0.056 | 0.048 |
| 0.4 | QR | **0.693** | **0.421** | 0.266 | 0.161 | 0.103 | 0.074 | 0.064 | 0.0563 | 0.053 | 0.05 |
| | PTC-A | 0.675 | 0.417 | 0.272 | 0.17 | 0.111 | 0.081 | 0.068 | 0.06 | 0.055 | 0.058 |
| | PTC-M | 0.667 | 0.416 | **0.274** | **0.177** | **0.113** | **0.081** | **0.068** | **0.061** | **0.057** | **0.057** |
| | TPOT-RL | 0.496 | 0.375 | 0.198 | 0.086 | 0.059 | 0.063 | 0.054 | 0.04 | 0.037 | 0.029 |
| 0.6 | QR | **0.632** | **0.326** | 0.185 | 0.103 | 0.062 | 0.045 | 0.037 | 0.03 | 0.029 | 0.027 |
| | PTC-A | 0.61 | 0.323 | 0.19 | 0.111 | 0.068 | 0.047 | 0.04 | 0.035 | 0.032 | 0.03 |
| | PTC-M | 0.596 | 0.322 | **0.194** | **0.113** | **0.069** | **0.049** | **0.039** | **0.035** | **0.032** | **0.03** |
| | TPOT-RL | 0.418 | 0.301 | 0.148 | 0.065 | 0.043 | 0.046 | 0.037 | 0.023 | 0.015 | 0.013 |

Table 6: Call success rates at various distances — topology of Figure 5(a)

| Load | Strategy | Min hop count | | | | | | |
|---|---|---|---|---|---|---|---|---|
| | | 1 | 2 | 3 | 4 | 5 | 6 | 7 |
| 0.1 | QR | **0.886** | **0.761** | 0.6 | 0.504 | 0.428 | 0.379 | 0.353 |
| | PTC-A | 0.876 | 0.746 | 0.586 | 0.495 | 0.426 | 0.381 | 0.365 |
| | PTC-M | 0.873 | 0.753 | **0.601** | **0.511** | **0.443** | **0.398** | **0.369** |
| | TPOT-RL | 0.187 | 0.129 | 0.11 | 0.102 | 0.104 | 0.139 | 0.203 |
| 0.2 | QR | **0.774** | **0.57** | 0.371 | 0.268 | 0.197 | 0.161 | 0.141 |
| | PTC-A | 0.77 | 0.561 | 0.369 | 0.265 | 0.2 | 0.163 | 0.145 |
| | PTC-M | 0.766 | 0.564 | **0.375** | **0.272** | **0.207** | **0.173** | **0.159** |
| | TPOT-RL | 0.214 | 0.127 | 0.085 | 0.065 | 0.055 | 0.06 | 0.061 |
| 0.4 | QR | **0.653** | 0.383 | 0.203 | 0.12 | 0.782 | 0.58 | 0.492 |
| | PTC-A | 0.648 | **0.385** | 0.208 | 0.126 | 0.828 | 0.631 | 0.53 |
| | PTC-M | 0.64 | 0.384 | **0.212** | **0.13** | **0.87** | **0.662** | **0.545** |
| | TPOT-RL | 0.202 | 0.109 | 0.054 | 0.025 | 0.021 | 0.024 | 0.048 |
| 0.6 | QR | **0.58** | 0.286 | 0.133 | 0.712 | 0.428 | 0.305 | 0.249 |
| | PTC-A | 0.566 | **0.289** | 0.14 | 0.781 | 0.481 | 0.347 | 0.287 |
| | PTC-M | 0.557 | 0.287 | **0.143** | **0.802** | **0.504** | **0.371** | **0.305** |
| | TPOT-RL | 0.177 | 0.095 | 0.046 | 0.017 | 0.011 | 0.0070 | 0.0050 |

Table 7: Call success rates at various distances — topology of Figure 5(b)

| Load | Strategy | Min hop count | | | | | | | | | | | |
|------|----------|---|---|---|---|---|---|---|---|---|---|---|---|
| | | 1 | 2 | 3 | 4 | 5 | 6 | 7 | 8 | 9 | 10 | 11 | 12 |
| 0.1 | QR | **0.894** | **0.755** | **0.616** | **0.452** | 0.324 | 0.247 | 0.208 | 0.175 | 0.149 | 0.12 | 0.087 | 0.082 |
| | PTC-A | 0.89 | 0.749 | 0.589 | 0.432 | 0.317 | 0.247 | 0.22 | 0.197 | 0.169 | 0.141 | 0.111 | 0.1 |
| | PTC-M | 0.88 | 0.728 | 0.598 | 0.443 | **0.336** | **0.267** | **0.238** | **0.223** | **0.191** | **0.164** | **0.124** | **0.122** |
| | TPOT-RL | 0.22 | 0.172 | 0.14 | 0.104 | 0.079 | 0.057 | 0.051 | 0.045 | 0.037 | 0.029 | 0.028 | 0.029 |
| 0.2 | QR | **0.847** | **0.647** | **0.47** | **0.296** | 0.189 | 0.123 | 0.098 | 0.079 | 0.061 | 0.047 | 0.036 | 0.029 |
| | PTC-A | 0.837 | 0.628 | 0.45 | 0.287 | 0.178 | 0.124 | 0.1 | 0.087 | 0.071 | 0.056 | 0.046 | 0.039 |
| | PTC-M | 0.823 | 0.622 | 0.448 | 0.292 | **0.192** | **0.136** | **0.109** | **0.096** | **0.086** | **0.066** | **0.053** | **0.046** |
| | TPOT-RL | 0.202 | 0.158 | 0.11 | 0.072 | 0.047 | 0.032 | 0.029 | 0.025 | 0.016 | 0.012 | 0.008 | 0.006 |
| 0.4 | QR | **0.77** | **0.502** | **0.315** | 0.17 | 0.094 | 0.057 | 0.041 | 0.031 | 0.022 | 0.017 | 0.011 | 0.012 |
| | PTC-A | 0.766 | 0.484 | 0.298 | 0.161 | 0.09 | 0.057 | 0.042 | 0.034 | 0.026 | 0.021 | 0.0145 | 0.015 |
| | PTC-M | 0.746 | 0.476 | 0.303 | **0.172** | **0.098** | **0.061** | **0.046** | **0.039** | **0.031** | **0.024** | **0.017** | **0.018** |
| | TPOT-RL | 0.189 | 0.141 | 0.09 | 0.056 | 0.034 | 0.02 | 0.0167 | 0.011 | 0.006 | 0.003 | 0.002 | 1.0E-4 |
| 0.6 | QR | **0.71** | **0.4** | **0.225** | 0.112 | 0.06 | 0.03 | 0.024 | 0.018 | 0.012 | 0.009 | 0.006 | 0.005 |
| | PTC-A | 0.706 | 0.393 | 0.216 | 0.109 | 0.057 | 0.031 | 0.025 | 0.0197 | 0.014 | 0.0117 | 0.008 | 0.0075 |
| | PTC-M | 0.688 | 0.386 | 0.221 | **0.113** | **0.061** | **0.036** | **0.028** | **0.023** | **0.017** | **0.013** | **0.009** | **0.008** |
| | TPOT-RL | 0.171 | 0.127 | 0.079 | 0.047 | 0.027 | 0.015 | 0.01 | 0.006 | 0.003 | 1.0E-4 | 1.0E-4 | 1.0E-4 |

In more detail, tables 5, 6, and 7 show the values of $x_d$ for different values of $d$ (the "min hop count"[22]) under different network loads for the topologies of figures 4, 5(a), and 5(b), respectively. These tables indicate the following trends. For very short distances (where the value of $d$ is less than 3), QR achieves a slightly better success rate than PTC-M. However, the success rate achieved at longer distances with either PTC-based strategies far outperform that of QR. Additionally, PTC performs better than TPOT-RL for all distances. To obtain a clearer picture of the relative advantage of the PTC-based strategies over QR and TPOT-RL in this context, we plot the relative improvement of $x_d$ that is achieved by using the PTC strategies over QR and TPOT-RL, $\frac{x_d^{PTC-Y} - x_d^{QR}}{x_d^{QR}}$ and $\frac{x_d^{PTC-Y} - x_d^{TPOTRL}}{x_d^{TPOT-RL}}$, respectively (where, $Y$ can be $A$ for the average capacity heuristic, or $M$ for the minimum capacity heuristic). Figure 11 shows the percentage change of $x_d$ achieved by using PTC-M over QR (Figure 11(a)), PTC-A over QR (Figure 11(b)), PTC-M over TPOT-RL (Figure 11(c)), and PTC-A over TPOT-RL (Figure 11(d)) at increasing values of the minimum hop count between call source and destination nodes in the irregular grid topology of Figure 4. Each plot in each of these figures is for a different value of the call origination probability. The graphs of Figure 12 and 13 show identical measures using the topologies of Figure 5(a) and 5(b), respectively.

Focusing on figures 11(a) and 11(b), it is observed that, for a given call origination probability, when the call destinations are very close to the call sources (i.e., when the "minimum hop count" axis has values of 1 and 2), QR performs slightly better than either PTC-M or PTC-A, indicated by the small negative deviation. For example, at load 0.6, QR achieves about a 5% improvement over PTC-M in connecting calls at nodes 1 hop away (see Figure 11(a)). However, with increasing distances between the call source and destination nodes

---

22. The maximum value of the minimum hop count is the property of the corresponding network.

(i.e., where the "minimum hop count" is 3 and above), the rate of successful connections is much higher for the PTC-based strategies than QR (the deviation values are positive). For example, in Figure 11(a), for a load value of 0.6, PTC-M achieves more than 15% improvement over QR in the connection rate of calls that have their destinations at least 8 hops away from the sources. Also, this relative improvement is observed to generally increase with increasing distance. Thus, the PTC information sharing strategies are more effective in connecting calls for which the source and destination nodes are farther apart. In figures 11(c) and 11(d), PTC-M and PTC-A perform better than TPOT-RL at all distances and under all load values.

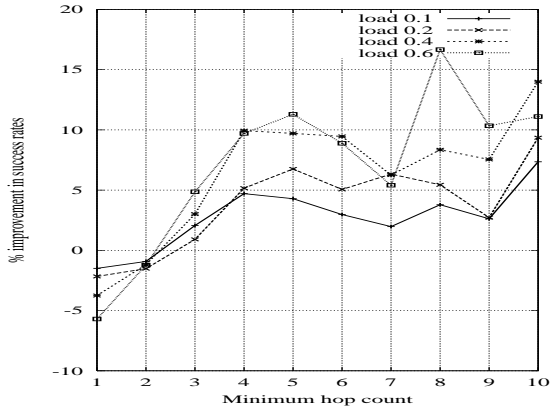

(a) PTC-M against QR

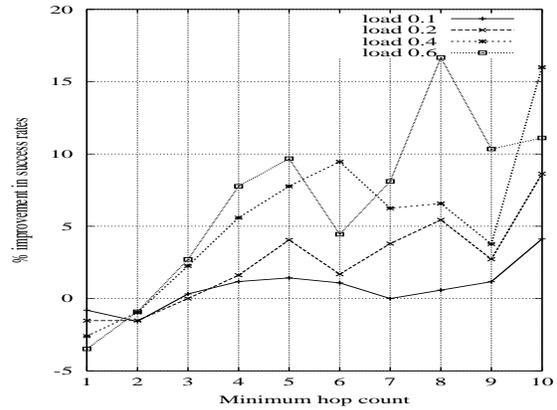

(b) PTC-A against QR

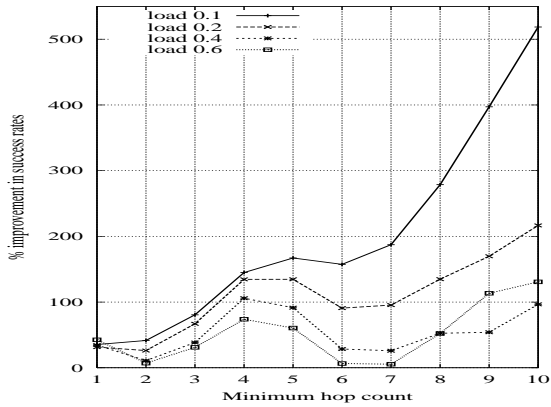

(c) PTC-M against TPOT-RL

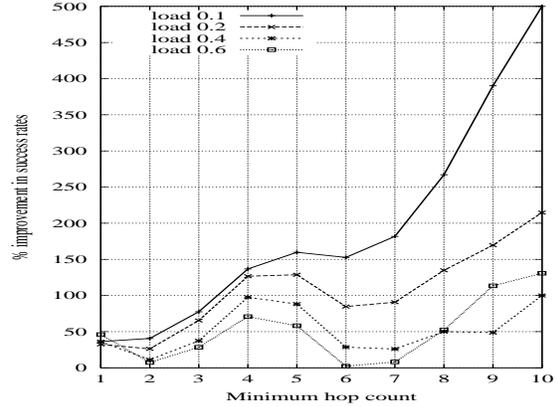

(d) PTC-A against TPOT-RL

Figure 11: Call success rates at various distances — topology of Figure 4

To explain this advantage of PTC, note that although the call connection process is executed in steps by multiple agents who use their individual estimates to forward a call, the forwarding decisions of the agents who are closer to the call origin are more critical in determining whether it will be successfully routed to the destination. This is because if these agents start forwarding a call in a direction where there is a high bandwidth occupancy on the

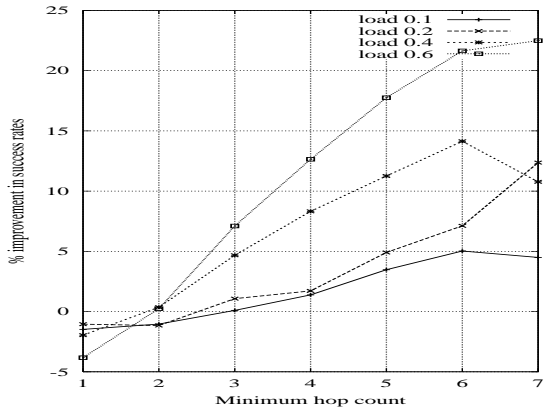

(a) PTC-M against QR

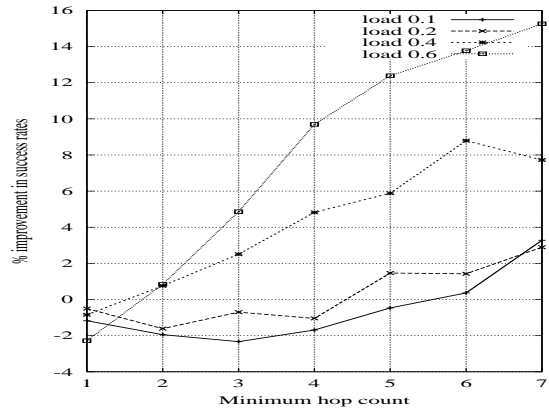

(b) PTC-A against QR

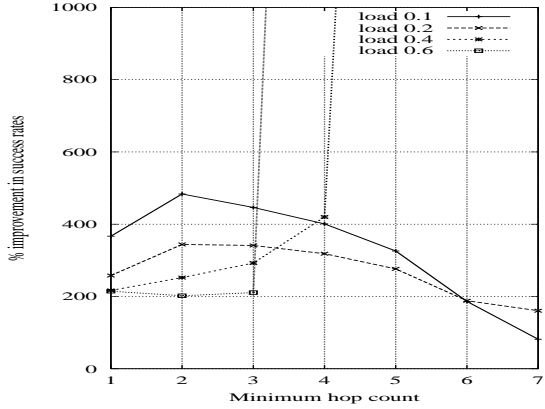

(c) PTC-M against TPOT-RL

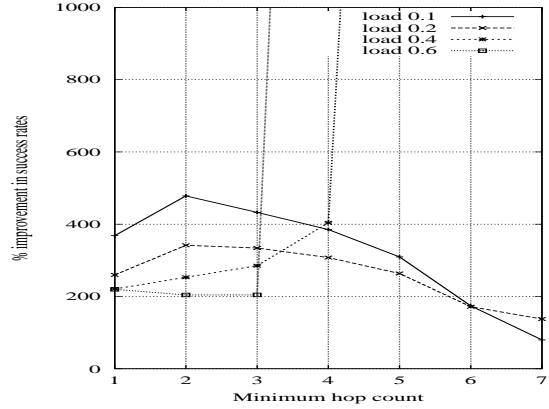

(d) PTC-A against TPOT-RL

Figure 12: Call success rates at various distances — topology of Figure 5(a)

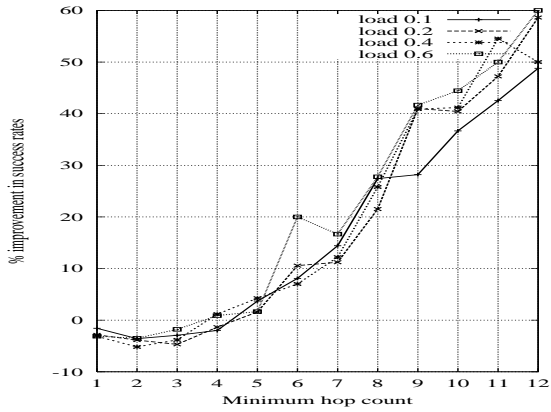

(a) PTC-M against QR

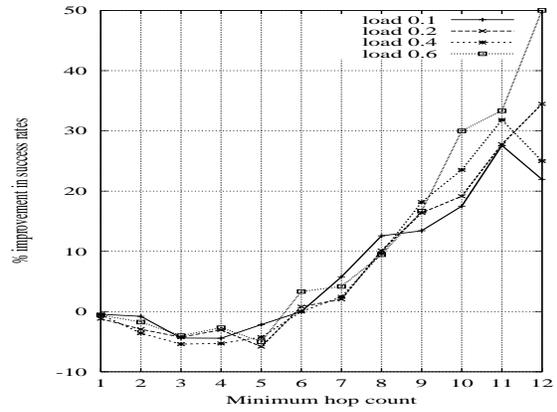

(b) PTC-A against QR

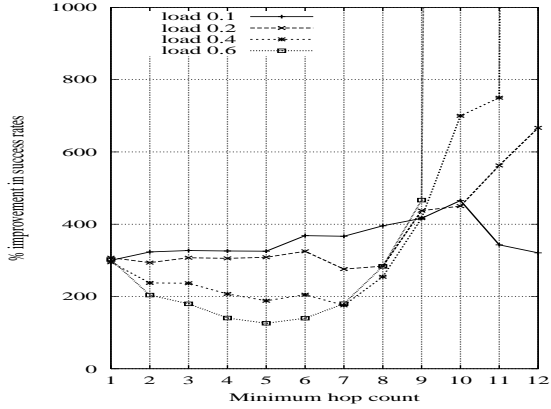

(c) PTC-M against TPOT-RL

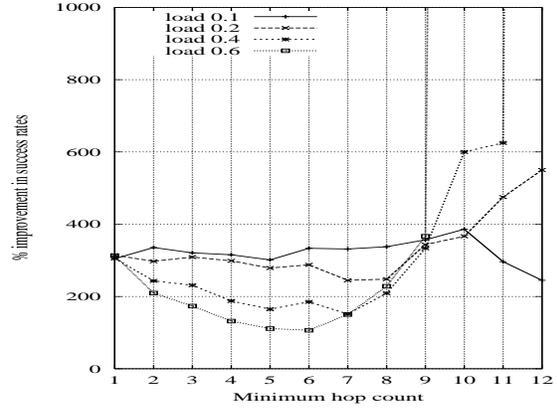

(d) PTC-A against TPOT-RL

Figure 13: Call success rates at various distances — topology of Figure 5(b)

nodes, then that will be a sub-optimal decision at the beginning of a task execution process. In this case, it is more likely that the call will be forwarded where there is no bandwidth left and, therefore, the forwarding process would terminate (the call would be dropped). With more up-to-date estimates achieved by PTC (as observed in the analysis of section 5), an agent is capable of taking better routing decisions in terms of forwarding in the appropriate direction than using QR. This is why we observe higher call success rates in the PTC-based strategies than QR when calls have to be routed at longer distances. Moreover, with increasing distance, successful routing becomes more difficult since the farther the destination the less up-to-date are the estimates. Given this, the increasing (more positive) deviation of the $x_d$ values with increasing distance, as observed in Figure 11, indicates that PTC is more capable (less affected) than either QR or TPOT-RL to route calls at long distances. PTC-M is more effective in placing long-distance calls than PTC-A although both are better than QR and TPOT-RL. This is evident by the slightly higher positive deviation of PTC-M over QR (Figure 11(a)) than that of PTC-A over QR (Figure 11(b)) or by the higher positive deviation of PTC-M over TPOT-RL (Figure 11(c)) than that of PTC-A over TPOT-RL (Figure 11(d)) for the same call origination probability.

The above observations from Figure 11 also hold in figures 12 and 13. Additionally, in all of these figures, it is observed that, for a given distance $d$, the deviation of the $x_d$ values are generally higher for a higher call origination probability. To justify this observation, note that the success rate of any given strategy decreases with increasing load (see section 7.2.1 for this result) because with more calls originating, the number of dropped calls increases since the nodes have only a limited amount of call channel bandwidth. Nevertheless, the communication strategy that generates better estimates would enable the agents to cope with the increasing load better by maintaining a higher call success rate. This is indicated by the observation that increased load impacts QR and TPOT-RL more than the PTC-based strategies, i.e., the decrease in the success rate is more in QR or TPOT-RL than in PTC-M or PTC-A with increasing load. Hence, the deviations of the $x_d$ values between PTC and the benchmarks increase with load. For example, in Figure 11(a), at a hop length of 8, the deviation of $x_d$ is about 2.5% with load 0.1, and more than 15% with load 0.6.

The observations in this section can be summarised as, *post-task-completion strategies are more capable of connecting calls at longer distances than the benchmark strategies and exhibit increased effectiveness in achieving this against high network loads*. A relative improvement of more than 50% is observed in the rate of successful call connections of PTC-M over QR at distances of 12 hops in a 100-node random graph with a high network load (see the graph for a load of 0.6 in Figure 13(a)). For the same parameter values, PTC-M achieves a large improvement over TPOT-RL of more than 1000% (see Figure 13(c)).

### 7.2.3 PERFORMANCE — INFORMATION MESSAGE RATE

The total number of messages (see section 7.1.3 for its definition) transmitted in the entire network is measured every T time steps of a simulation. The average number of such messages per unit time is computed over the steady state. These measurements are repeated for each of the three topologies. Results from constant-load experiments are reported first followed by those from a dynamically changing load.

**Constant Load.** Table 8 shows the message rate obtained in each of the strategies under steady state conditions in the topology of Figure 4. Tables 9 and 10 show the same for the topologies of Figure 5(a) and Figure 5(b), respectively. In all of these results, it is observed that for any given call probability, the message rate (under column "Avg") is significantly lower in the PTC strategies than both QR and TPOT-RL. For example, in Table 8, it is 0.25 for PTC-M, 0.26 in PTC-A, 0.39 for QR, and 0.523 for TPOT-RL with a load level of 0.1.

An additional observation is that the relative reduction of the message rate (column named "% Saving" shows these values) achieved by using PTC becomes more pronounced with increasing load. These values are computed as $\frac{|m^{PTC-Y}-m^{QR}|}{m^{QR}}$, $\frac{|m^{PTC-Y}-m^{TPOT-RL}|}{m^{TPOT-RL}}$, where $m$ represents the average steady state message rate of a given communication strategy for a given load (and $Y$ can be either $A$ or $M$). For example, in Table 8, at load 0.1, PTC-M has about 35.6% $(=|\frac{0.251-0.39}{0.39}|)$ less rate of messages than QR and about 52% $(=|\frac{0.251-0.523}{0.523}|)$ less than TPOT-RL. However, this saving in the message rate increases to about 80.3% $(=|\frac{0.285-1.45}{1.45}|)$ relative to QR and to about 72.8% $(=|\frac{0.285-1.05}{1.05}|)$ relative to TPOT-RL when the call probability is 0.6. This is because with increasing network load, the increase in the number of messages in both QR and TPOT-RL is much higher than the increase in the PTC-based strategies (e.g., in Table 8, the message rate increases from 0.39 to 1.45 in QR — a 272% increase, from 0.523 to 1.05 in TPOT-RL — a 101% increase, and from 0.251 to 0.285 in PTC-M — only a 13.5% increase — as the load increases from 0.1 to 0.6).

Table 8: Information message rates for all strategies — topology of Figure 4

| Load | Strategies | | | | | | | | % Saving | | | |
|------|------|------|------|------|------|------|------|------|------|------|------|------|
| | QR | | PTC-A | | PTC-M | | TPOT-RL | | PTC-A / | PTC-M / | PTC-A / | PTC-M / |
| | Avg | Stdev | Avg | Stdev | Avg | Stdev | Avg | Stdev | QR | QR | TPOT-RL | TPOT-RL |
| 0.1 | 0.39 | 0.0048 | 0.26 | 0.0019 | **0.251** | 0.0019 | 0.523 | 0.0023 | 33.33 | 35.64 | 50.28 | 52.00 |
| 0.2 | 0.66 | 0.0079 | 0.281 | 0.002 | **0.275** | 0.0022 | 0.683 | 0.0029 | 57.42 | 58.33 | 58.85 | 59.73 |
| 0.4 | 1.1 | 0.01 | 0.289 | 0.0018 | **0.284** | 0.0016 | 0.891 | 0.0036 | 73.73 | 74.18 | 67.56 | 68.12 |
| 0.6 | 1.45 | 0.0125 | 0.2894 | 0.0017 | **0.285** | 0.0015 | 1.05 | 0.0077 | 80.04 | 80.34 | 72.43 | 72.85 |

Table 9: Information message rates for all strategies — topology of Figure 5(a)

| Load | Strategies | | | | | | | | % Saving | | | |
|------|------|------|------|------|------|------|------|------|------|------|------|------|
| | QR | | PTC-A | | PTC-M | | TPOT-RL | | PTC-A / | PTC-M / | PTC-A / | PTC-M / |
| | Avg | Stdev | Avg | Stdev | Avg | Stdev | Avg | Stdev | QR | QR | TPOT-RL | TPOT-RL |
| 0.1 | 0.33 | 0.0023 | 0.23 | 0.0012 | **0.23** | 0.0014 | 0.759 | 0.0054 | 30.30 | 30.30 | 69.69 | 69.69 |
| 0.2 | 0.55 | 0.0048 | 0.26 | 0.0013 | **0.25** | 0.0012 | 1.42 | 0.0074 | 52.73 | 54.55 | 81.69 | 82.39 |
| 0.4 | 0.9 | 0.0049 | 0.27 | 0.001 | **0.26** | 0.0009 | 2.163 | 0.0091 | 70.0 | 71.11 | 87.51 | 87.97 |
| 0.6 | 1.2 | 0.0054 | 0.272 | 0.0007 | **0.267** | 0.0012 | 2.70 | 0.0106 | 77.33 | 77.75 | 89.92 | 90.11 |

Note that in QR, a path agent transmits a new information message to the forwarding (upstream) agent at each step of the call forwarding process. Thus, with increased load, this strategy incurs a large increase in the number of messages because there are many more calls to be routed. In TPOT-RL, a node forwards its state information while routing a call. So, although reward distribution occurs in TPOT-RL every `update-interval` time steps, the state information messages propagated during call routing contribute towards the message overhead incurred in the system. So, when larger numbers of calls need to be routed at high load values, the number of such messages increases. The PTC-based strategies, on the other hand, attain a significant saving in the message rate by delaying the information transmission until a call connects and then transmitting a single message from the destination to the source node (with only updating the reward information at each step of the message propagation). This prevents any information exchange from occurring for calls that fail to connect. Both QR and TPOT-RL, on the other hand, would still incur the messaging cost even if a call finally fails to connect. In addition, the PTC-based strategies save some messages from being exchanged that QR and TPOT-RL incur on the loops of a call route. Because the agents using PTC transmit messages only after a call is connected, and the loops on call routes are dropped (as a call is forwarded), messages are prevented from being transmitted on loop portions.

Table 10: Information message rates for all strategies — topology of Figure 5(b)

| Load | Strategies | | | | | | | | % Saving | | | |
|---|---|---|---|---|---|---|---|---|---|---|---|---|
| | QR | | PTC-A | | PTC-M | | TPOT-RL | | PTC-A / QR | PTC-M / QR | PTC-A / TPOT-RL | PTC-M / TPOT-RL |
| | Avg | Stdev | Avg | Stdev | Avg | Stdev | Avg | Stdev | | | | |
| 0.1 | 0.58 | 0.0117 | 0.214 | 0.0009 | **0.21** | 0.0014 | 1.42 | 0.018 | 63.10 | 63.79 | 84.92 | 85.21 |
| 0.2 | 0.83 | 0.0112 | 0.235 | 0.001 | **0.229** | 0.001 | 2.19 | 0.0091 | 71.69 | 72.41 | 89.27 | 89.54 |
| 0.4 | 1.26 | 0.0098 | 0.249 | 0.001 | **0.241** | 0.0008 | 2.98 | 0.0061 | 80.24 | 80.87 | 91.64 | 91.91 |
| 0.6 | 1.63 | 0.0095 | 0.253 | 0.0007 | **0.245** | 0.0008 | 3.37 | 0.0079 | 84.48 | 84.97 | 92.49 | 92.72 |

**Dynamically Changing Load.** Similar to the call success rate measurements of section 7.2.1, the time-variation of the information message rates are recorded for all strategies when the network load fluctuates dynamically in a simulation run.

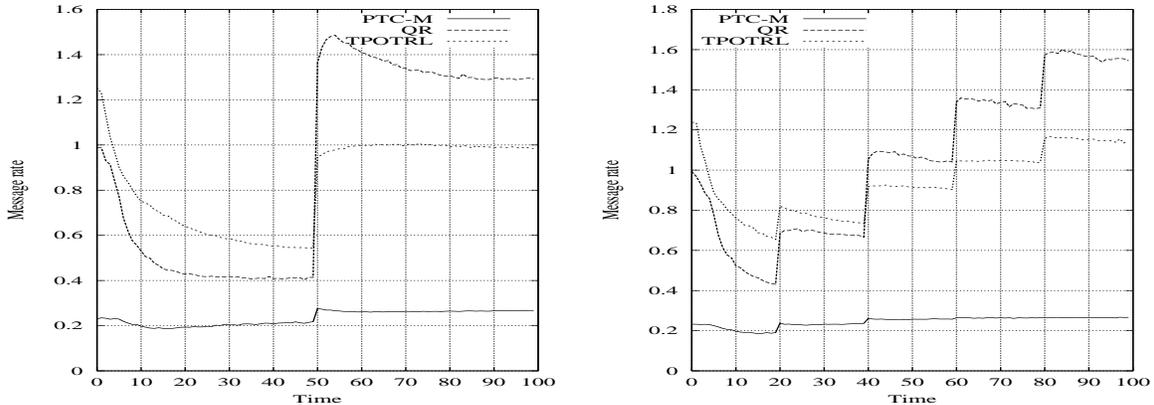

(a) With 2 degrees of dynamism

(b) With 5 degrees of dynamism

Figure 14: Time variation of message rates of QR, PTC-M, and TPOT-RL with network load fluctuations — topology of Figure 4

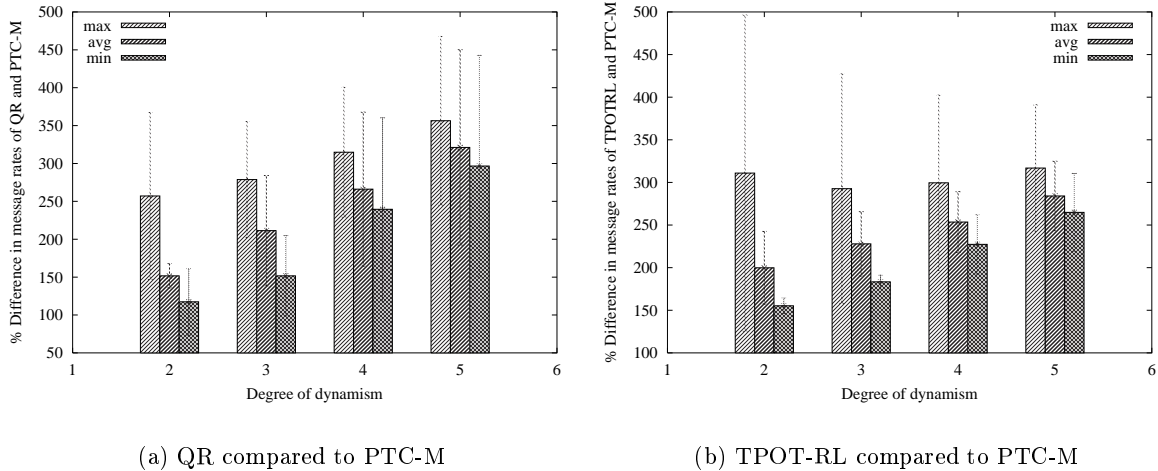

(a) QR compared to PTC-M  (b) TPOT-RL compared to PTC-M

Figure 15: Summary statistics of message rate differences between QR, TPOT-RL, and PTC-M with network load fluctuations — topology of Figure 4

Figure 14(a) shows the message rate variations of PTC-M, QR, and TPOT-RL when the load level changes from 0.1 to 0.6 in a simulation run using the topology of Figure 4. Figure 14(b) shows the same when the load is varied as 0.1, 0.2, 0.4, 0.6, and 0.8 after equal intervals of time in the course of a simulation run. Both of these figures show that PTC-M has a significantly lower message rate than both QR and TPOT-RL under fluctuating load conditions. Further, as the network load increases dynamically both QR and TPOT-RL incur a large increase in the message rates while the increase in PTC is insignificant. In both QR and TPOT-RL, information is transmitted during call setup (although reward updates occur in TPOT-RL after every `update-interval` time steps). Hence, as more calls originate with increasing network load, the number of such information propagations increase in these strategies. It is also observed that with increasing load, TPOT-RL has a lower message rate than QR. We identify that this relative advantage of TPOT-RL against QR is due to its poorer call success rate than QR (reported in section 7.2.1). Since TPOT-RL is less efficient than QR in connecting calls (this implies that in TPOT-RL a large number of call attempts are unsuccessful), and since both propagate information while call forwarding, TPOT-RL does not incur as much of an increase in the number of messages as incurred by QR with increasing load.

In a way similar to the success rate results with fluctuating load, the summary of the differences in message rates between QR and PTC-M and between TPOT-RL and PTC-M are presented in figures 15(a) and 15(b), respectively. In both of these figures, the degree of dynamism is plotted along the horizontal axis while the percentage increase of message rate using QR or TPOT-RL against PTC-M is plotted along the vertical axis. As observed in the success rate results of Figure 10, in both Figure 15(a) and 15(b), the gap between the minimum and the maximum differences of message rates reduce with increasing degrees of freedom. For instance, in Figure 15(a), this gap is about 140% with 2 degrees of freedom, while it is 60% with 5 degrees of freedom. Nevertheless, the mean difference between the message rates of PTC-M and QR or that between PTC-M and TPOT-RL increases with increasing degrees of dynamism. For example, the mean difference increases from 151% to 321% in Figure 15(a), while it increases from 200% to 284% in Figure 15(b). This indicates the advantage of PTC in terms of maintaining a limited number of message overhead compared to both QR and TPOT-RL.

From these observations, it can be concluded that the *post-task-completion strategies not only achieve a higher call success rate (section 7.2.1) and higher effectiveness in connecting calls at longer distances under high loads (section 7.2.2) than the benchmark strategies, but also achieve these at the expense of a significantly lower rate of messages under both static and dynamically changing load conditions.* For example, PTC-M achieves an 80% saving in message rate compared to QR, and a 72% saving compared to TPOT-RL in the grid topology under high network load. Further, in a dynamic load setting, with five changes in the network load, PTC-M saves about 320% in message rate than QR and about 284% than TPOT-RL. Thus, PTC is shown to be a *more efficient strategy towards developing a cooperative MAS for the distributed resource allocation problem.*

## 8. Conclusions and Future Work

This paper focuses on the critical issue of allowing agents with partial observability to effectively cooperate on complex tasks that require the participation of multiple agents for successful completion. To achieve this goal, it is required that the agents take their actions based on estimates of the states of other agents so that a task is solved in a consistent manner. Nevertheless, without being able to directly observe the system states, it is difficult for the agents to develop such estimates. A further difficulty in achieving the objective is that most systems are dynamic, therefore, the system states are subjected to continuous changes. Hence, in such scenarios, the agents should have a way of adapting their estimates with such changes.

Given this, we show that cooperative information sharing is a practical and effective mechanism to allow the agents to estimate the unobserved states. Coupled with this mechanism, we use Q-learning to produce robust and flexible estimations of these states. In particular, the specific contribution of this paper has been the introduction of a novel information distribution principle — post-task-completion information sharing — to improve the learning of the state estimates. In comparison to the protocol of learning from nearest neighbours' information, we have established, using formal analysis, that our strategy is capable of maintaining more up-to-date estimates by ensuring a more timely distribution of information. Further, we have implemented a set of communication heuristics based on the PTC-sharing principle to be used in a practical multi-agent resource-allocation problem (call routing in a telephone network). We have conducted extensive empirical studies by comparing our strategy against a set of benchmark algorithms across a wide range of environmental settings by selecting different network topologies, network loads, and dynamically changing load patterns. Results from these studies have shown that, compared to the benchmarks, our protocol achieves a higher call success rate (up to 60%) and a superior performance (of more than 1000%) in the ability to successfully route calls to long distances. It has been also demonstrated that our strategy attains these improvements at a significantly lower message overhead (up to 300%) than the benchmarks.

We emphasise that the PTC-sharing is a *generic* mechanism to improve the learning of state estimates by distributed agents. This is observed in the analysis of section 5, where no domain-specific assumptions were made. However, specific communication heuristics, such as PTC-M and PTC-A in this paper, can be designed around this mechanism for a given problem. Thus, we envisage that our strategy is applicable to a wide range of task domains as discussed in section 4. To verify our hypothesis, we are employing a PTC-based heuristic to perform distributed fault detection in a network and are obtaining promising initial results.

An additional benefit of our information-sharing strategy is that it is *practically applicable* since it is developed around such realistic assumptions as local observability, actions restricted within the local state space, dynamic environment, finite time delay for a task to be completed by multiple agents in sequence, and latency involved in the propagation of information between agents. Moreover, our communication strategy is a contribution to machine learning research in general since it is a practical and effective means of improving learning in real-world MAS. In this context, we note that a key challenge in machine learning research is to develop algorithms for sequential decision making in uncertain domains (Russel & Norvig, 2002). A popular way of solving this problem is to formulate a decentralised Markov Decision Process (DEC-MDP) (Becker, Zilberstein, Lesser, & Goldman, 2003; Bernstein, Givan, Immerman, & Zilberstein, 2002) to provide the necessary control to the agents to take actions against the dynamics of the domain. In this context, we believe that our communication strategy can contribute significantly by generating high-fidelity solutions to a DEC-MDP that models a practical dynamic MAS. Building on this idea, we plan to investigate ways of integrating our communication strategy into such a model and study the impact on the quality of solutions generated.

## Acknowledgments

Our sincere thanks to the reviewers whose informative and insightful comments have enabled us to improve the paper substantially. We thank Xudong Luo, Steve Gunn, Srinandan Dasmahapatra, and Steve Braithwaite for their comments and suggestions. Finally, we are obliged to EPSRC for funding the project *MO*bile *H*andsets *I*n *C*ooperative *A*gents *N*etworks (MOHICAN), GR/R32697/01, of which this work is a part.